\documentclass[aps,prb,twocolumn,showpacs,superscriptaddress]{revtex4}
\usepackage{bm,amsmath,amssymb,latexsym,mathrsfs,enumerate}
\usepackage[mathcal]{euscript}
\usepackage[cp1251]{inputenc}\usepackage[english,russian]{babel}
\usepackage{graphicx, subfigure}
\newcommand{\figwidth}{0.44\textwidth}
\newcommand{\figwidths}{0.2\textwidth}

\begin{document}

\title{Ground states and magnetization process for an triangular lattice
array of magnetic dots with perpendicular anisotropy}

\author{V.~E. Kireev}
\affiliation{Institute of Magnetism, 03142 Kiev, Ukraine}

\author{R.~S. Khymyn}
\affiliation{Institute of Magnetism, 03142 Kiev, Ukraine}

\author{B.~A. Ivanov}
\email{bivanov@i.com.ua} \affiliation{Institute of Magnetism, 03142
Kiev, Ukraine} \affiliation{National Taras Shevchenko University of
Kiev, 03127 Kiev, Ukraine}

\author{C.~E. Zaspel }
\affiliation{University of Montana-Western, Dillon, Montana 59725,
USA. }

\date\today

\begin{abstract}
We analyzed the ground state of the array of magnetic particles
(magnetic dots) which form a two-dimensional triangular lattice, and
magnetic moment of which is perpendicular to the plane of the
lattice, in the presence of external magnetic field. In the small
fields long range dipole-dipole interaction leads to the specific
antiferromagnetic order, where two out of six nearest neighbors of
the particle have the same direction of magnetization moment and
four - the opposite one. It is shown that magnetization process in
such array of particles as opposed to the rectangular lattices
results from the formation of the magnetized topological defects
(dislocations) in the shape of the domain walls.
\end{abstract}



\maketitle

\section{Introduction.}

Magnetic ordering is usually attributed with exchange interaction of
atomic spins, leading to rather simple magnetically ordered
states.~\cite{SW,andreev80} Long-range magnetic dipole interaction
usually produces smooth non-uniformity (domain structures of
different kind) above this simple exchange
structure.~\cite{Hubert,BarYablUFN,BarIvJETP77} Nevertheless,
systems of magnetic moments with \emph{pure} dipolar interaction,
so-called dipolar magnets, have been theoretically studied for more
than sixty years,\cite{LutTisza} and many physical properties,
lacking in the spin-exchanged systems, are known for those models.
Note first the presence of a non-unique ground state with nontrivial
continuous degeneracy for quite simple bipartite lattices, like
three-dimensional cubic lattice,\cite{LutTisza,BelobGIgnat} and for
two-dimensional square lattice,\cite{DipD2,Gus99,dipObzorUFN} as
well as specific phase transitions induced by external magnetic
field.\cite{BishGalkIv,GalkinIvMerk,GalkinIvSF} Magnon spectra for
dipolar magnets demonstrate non-analytic behavior either for small
wavevectors,\cite{GalkinIvZasp06,BondGIvZasp,Galkin+CollModJM3} or
at some symmetrical points within the Brillouin zone.
\cite{BondGIvZasp} The Mermin–Wagner theorem is not valid for
two-dimensional magnets with a dipolar coupling of spins with
continuous degeneracy, and a true long range order can exist even
for a purely two-dimensional case at finite
temperatures.\cite{Maleev,Bruno,IvTartPRL}

The models of dipolar magnets were discussed originally in regard to
real crystalline spin systems. During the last decade the most
impressive achievements in magnetism were related to fabrication,
investigation, and application of artificial magnetic materials, see
Ref.~\onlinecite{Skomski} for a recent review. The nanotechnologies
today have progressed to the state where the manufacture of
nanosize, periodic magnetic superlattices of different types is
feasible. Among them two-dimensional lattices of sub-micron magnetic
particles (so-called magnetic dots) attract much attention. These
magnetic dots, of different forms and of a submicron size, are made
of soft magnetic materials such as Fe, Ni, Co and
permalloy,\cite{Hillebrands,Miramond,Wassermann,Runge,Cowburn} or
highly anisotropic materials like dysprosium,\cite{WeissKlitzing} or
FePt.\cite{FePt}

In the dot array lattice dots are separated from each other so that
direct exchange interaction between the dots is negligible. Thus the
dipolar interaction is the sole source of coupling between dots and
the configuration of dot magnetic moments is dictated by the dipolar
interactions of the dots and by the external field. Owing to the
absence of exchange, magnetic dot arrays constitute promising
material for high-density magnetic storage media. For this purposes,
the dense arrays of small enough magnetic dots with the magnetic
moments perpendicular to the array plane are optimal, see
Refs.~\onlinecite{chou94,Meier98,Ross}. Currently, ordered arrays of
magnetic submicron elements have been discussed as materials for
so-called \emph{magnonics}, i.e., a new field in the applied physics
of magnetism in which magnon modes with a discrete spectrum present
for magnetic nanoelements are used in devices for processing
microwave signals.\cite{Magnonics}

For small enough dots with size of the order of 100 nm the
magnetization inside of a dot is almost uniform, producing the total
magnetic moment $m_0 \gg\mu _B $, where $\mu _B $ is the Bohr
magneton, the typical value for an atomic magnetic moment. For
rather small magnetic dots of volume $10^4-10^6$ nm$^3$ the value of
$m_0$ exceeds $10^4\mu _B $, and for dense arrays the characteristic
energy is higher than the energy of thermal motion at room
temperature.\cite{BishGalkIv,GalkinIvMerk} The individual dots in an
array do not touch each other, and their interaction is only
determined by the dipole interaction of the magnetic moments.
Therefore, magnetic dot arrays represent a new kind of magnetic
material with purely two-dimensional lattice structure and high
enough pure dipolar coupling between magnetic moments. Such systems
represent dipolar magnets and fill their theoretical investigation
with a new physical content. Thus, magnetic dot arrays are
interesting as radically new objects for the fundamental physics of
magnetism.

We will discuss only the situation where the magnetic moment of an
individual magnetic dot is perpendicular to the array plane
($xy$-plane), $\mathbf{m} = \pm m_0\mathbf{e}z$, and the system can
be described on the basis of the Ising model. This situation is most
promising for perpendicular magnetic recording systems. The energy
of dipolar interaction of Ising moments perpendicular to the
system's plane is minimal for antiparallel orientation of magnetic
moments, that, of course, can not be fulfilled for \emph{any} pair
of particles in the array. Within the nearest-neighbors
approximation, such interactions lead to antiferromagnetic (AFM)
structures, e.g., simple chessboard AFM ordering is known for
two-dimensional Ising square lattice with dipolar
interaction.\cite{BishGalkIv} To explain the validity of this result
for real dipolar interaction it is enough to mention that accounting
for next nearest neighbors gives a correction to the energy of the
order of 30 \% only.

Thus, the magnetic structure for square lattice dot arrays can be
easily understood. However, the close-packed triangular lattices of
the magnetic dots are also frequently used in experiments. In
particular, these lattices of cylindrical particles considerably
extended in the direction normal to the array plane are naturally
obtained when the array is prepared by controlled
self-organization.\cite{TreugMass} Again, the properties of these
systems can be described in the two-dimensional Ising model with AFM
interactions. However the triangular lattice with AFM interaction of
the moments is a typical example of frustrated antiferromagnets, see
for review.\cite{GekhtObz}

For frustrated magnets spins interact through competing exchange
interactions that cannot be simultaneously satisfied, giving rise to
a large degeneracy of the ground state of the system. For
nearest-neighbor Ising triangular lattice with AFM interaction, the
thermodynamic properties are quite unusual.\cite{GekhtObz} It is
enough to mention that in this model magnetic ordering is absent at
any finite temperature $T \neq 0$; the ordering appears as a result
of accounting for the next-nearest-neighbor interactions
only.\cite{NoPT} This counterintuitive feature can be explained
within the concept of creation of linear topological
defects.\cite{Korshunov} Thus, in contrast to bipartite square
lattice, nearest-neighbors approximation did not provide even
adequate zero approximation to the problem of the ground state of a
triangular lattice.

In the present work the ground state of a triangular lattice of
mesoscopic magnetic dots, each having a strong easy axis for
magnetization perpendicular to the array plane, and in an external
magnetic field also perpendicular to the plane of the dot lattice,
will be considered. Both unbounded planar lattice and various
bordered semi-infinite or finite elements of the triangular lattice
are investigated. A cascade of phases with different patterns of dot
magnetization has been found; these constitute the sequence of
ground states as a function of the external magnetic field. In
contrast to a square lattice, the transition between these states is
governed by a novel mechanism involving creation of linear
topological defects with non-zero
magnetization.\cite{IvanovKireev09}

\section{Model description.}
Consider the set of Ising magnetic moments $\mathbf{m}_n = \sigma
 m_0 \mathbf{e}_z $, $\sigma
\pm 1$, each parallel to the $z$-axis, and placed in a sites of a
triangular lattice $\mathbf{n}$,
\begin{equation}\label{D6z:W}
\mathbf{n}=ak\mathbf{e}_x +
\frac{al}{2}\left(\mathbf{e}_x+\sqrt{3}\mathbf{e}_y\right).
\end{equation}
where $a$ is a lattice constant, $k, l$ are integers, and
$\mathbf{e}_x$ and $\mathbf{e}_y$ are unit vectors parallel to $x$
and $y$ axis, respectively. The magnetic moments are interacting
through the magnetic dipole interaction, and an external magnetic
field $\mathbf{H}=H\mathbf{e}_z$ is applied perpendicularly to the
array's plane. The Hamiltonian of this system of magnetic moments
can be written as
\begin{equation}\label{D6z:W}
W = m_0^2 \sum_{\mathbf{n} \neq \mathbf{n}'} \frac{\sigma_\mathbf{n}
\sigma_{\mathbf{n}'}}{|\mathbf{n}-\mathbf{n}'|^{3}} - m_0 H
\sum_\mathbf{n} \sigma_\mathbf{n} \;,
\end{equation}
where the first term describes dipolar interaction, with the
summation performed over all of the pairs of the lattice sites.
Below for the sake of simplicity we will present the energy (per one
magnetic particle) in the units of $m_0^2 / a^3$ and we will use the
dimensionless magnetic field, $h=H/H_{*}$, where characteristic
value $H_{*} = m_0 / a^3$. The present model is clearly not
restricted to a dot lattice of the type explicitly described above,
but also applies directly to any triangular lattice of identical
dipoles that are restricted to the two directions of normal
orientation.\cite{dipObzorUFN}  It is interesting that the model
formulated in this paper can be used to describe a system of a
vortex state magnetic dots, accounting for the interaction of a
magnetic moment of vortex cores.\cite{BishGalkIv}

As has been mentioned above, the triangular lattice is known as a
typical frustrated lattice for antiferromagnetic ordering, and the
frustration is present even for simplest nearest--neighbor
interaction.  Of course, not only nearest neighbors are important
for magnetic dipole interaction. The dipole-dipole interaction is
long ranged, that frequently leads to quite complicated structures
with many sublattices and with high level of degeneracy. The
presence of these two sources of degeneracy makes the problem less
definite, and it is not obvious \emph{a priori} what structure will
constitute the ground state for such a lattice. In this situation it
is natural to start with the numerical analysis of the problem.

\section{GROUND STATES for infinite system: NUMERICAL ANALYSIS}

We perform Monte-Carlo analysis (simulated annealing, see for
details Appendix) of the magnetic configurations with minimal energy
at zero magnetic field as well as for for different values of the
magnetic field.

\subsection{Simplest ground states of the system: zero field and saturation. }

It is clear that for high enough magnetic field all magnetic moments
will be parallel to the field, giving the saturated magnetic
structure, which can be referred to as a ferromagnetic structure.
The ground state of the array in the absence of magnetic field is
much less trivial.

\begin{figure}[h]
\begin{center}
\includegraphics[bb = 174  618 442  697, width=0.47\textwidth]{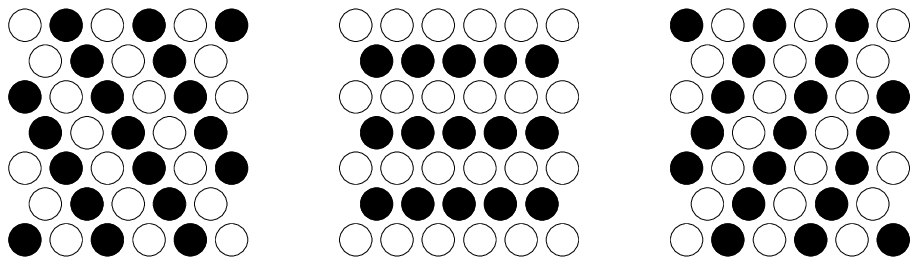}
\end{center}
\caption{Uniform states of the 4–2 type giving the energy minimum at
small magnetic field. Here and below at all figures the open and
closed circles denote the particles with
the upward and downward moments, respectively.} %
\label{f:2-4states} %
\end{figure}

It is known that for triangular Ising lattice in the
several-neighbor approximation the simple AFM order with two
sublattices can be implemented.\cite{SlotteHemmer84} As well, we
found the same configuration  for the long-range dipolar interaction
in the absence of the field and for a small enough magnetic field.
For these states, the magnetic elementary cell is rectangular having
lower symmetry than for the underlying triangular lattice, and this
state possesses much higher discrete degeneracy than the simple
chessboard structure for a square lattice discussed before. Several
such AFM states can occur in the system, which are different but
fully equivalent in their energies. Fig.~\ref{f:2-4states} presents
three of these states, while the other three states are obtained
from them by changing the magnetic moment sign $\sigma_{\mathbf{n}}$
at all of the particles. Then, we will briefly describe the magnetic
state of a given particle, specifying the number of neighbors with
the favorable and unfavorable orientations. The antiferromagnetic
states present on the figure \ref{f:2-4states} can be called the 4–2
type state. Of all of the states of the system with the simple
periodic distribution of the moments, this 4–2 type state
corresponds to the \emph{minimal} energy. It is not \emph{optimal}
because only four neighbors of the six nearest neighbors of each
moment have the sign opposite to the sign of this moment, while the
remaining two neighbors are parallel, and this is unfavorable to the
AFM interaction. The necessity of deviation from the 6–0 optimal
structure is a typical manifestation of frustration in the system.

\subsection{Monte-Carlo analysis for intermediate field values.}
\begin{figure}
\includegraphics[width=\figwidth]{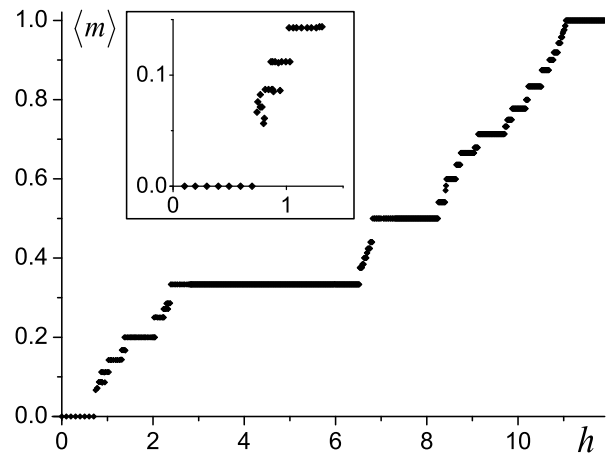}
\caption{Mean value of the magnetization $\langle m_0 \rangle$ (in
units of $m_0$, per one dot) of the array as a function of magnetic
field (in units of $H_{*}=M_0/a^3$) found by Monte-Carlo
simulations. Detailed data at low fields are present at the insert.
\label{f:D6z:magn_inf}}
\end{figure}

As was mentioned above, for a zero magnetic field the simple AFM
structure with zero mean value of the magnetic moment is present.
Monte-Carlo analysis shows that for some small but finite values of
the magnetic field, at least up to $h = 0.7$, the mean value of the
magnetic moment $\langle m \rangle $ equals zero indicating to the
simple AFM structure. For higher fields, numerous more complex
structures with $0 < \langle m \rangle < m_0$ occur in the
intermediate region between AFM state and saturated state. The mean
value of the magnetic moment (per one particle) $\langle m \rangle$
corresponding to these configurations, is present on the
Fig.~\ref{f:D6z:magn_inf}.

\begin{figure}
 \subfigure[\ $h = 1.0$
\label{f:D6z:inf_p9}] {\includegraphics*[bb = 20 500 570 840,
width=\figwidths]{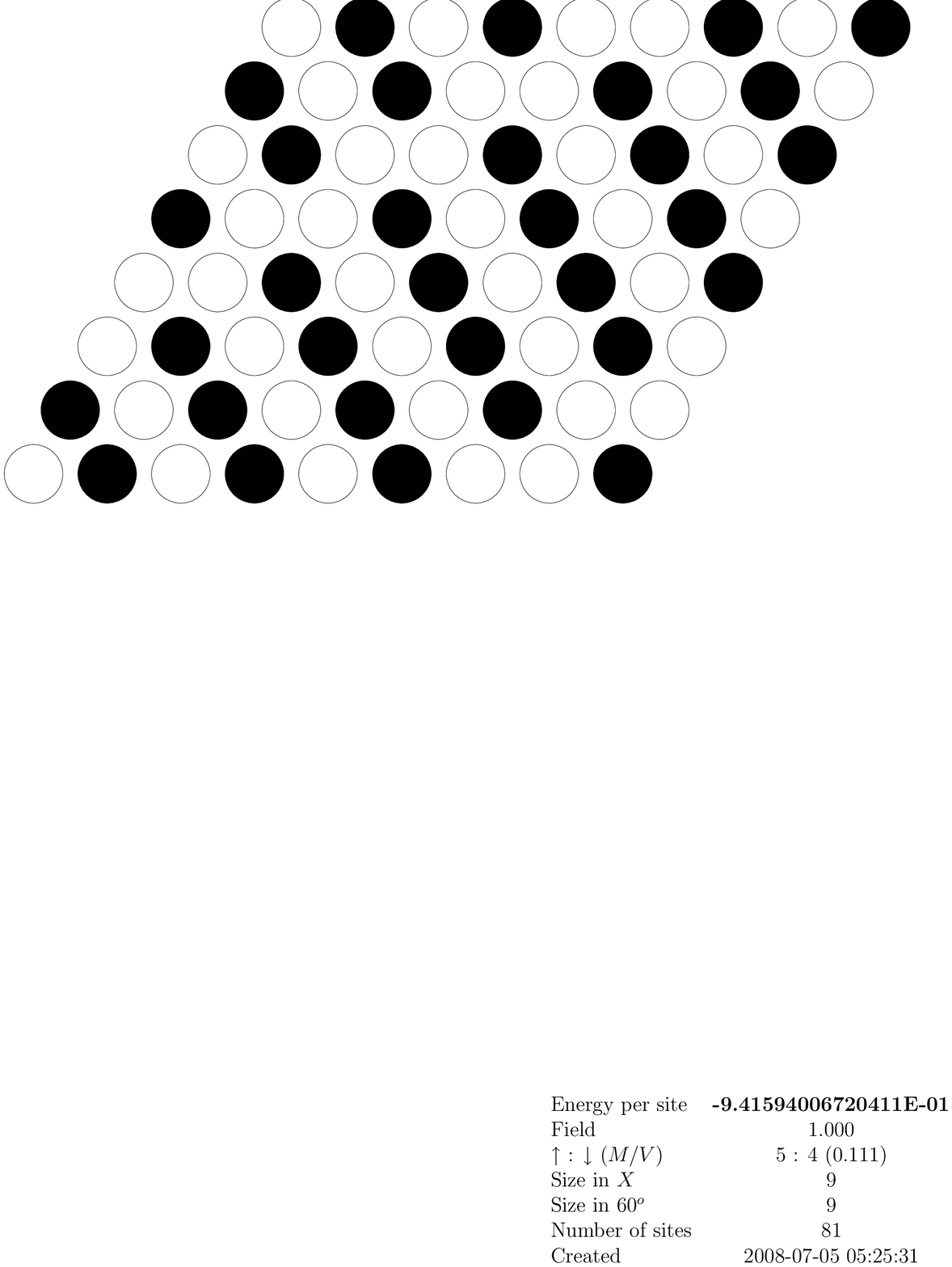}}\hfill \subfigure[\ $h = 1.1$
\label{f:D6z:inf_p7}] {\includegraphics*[bb = 20 500 570 840,
width=\figwidths]{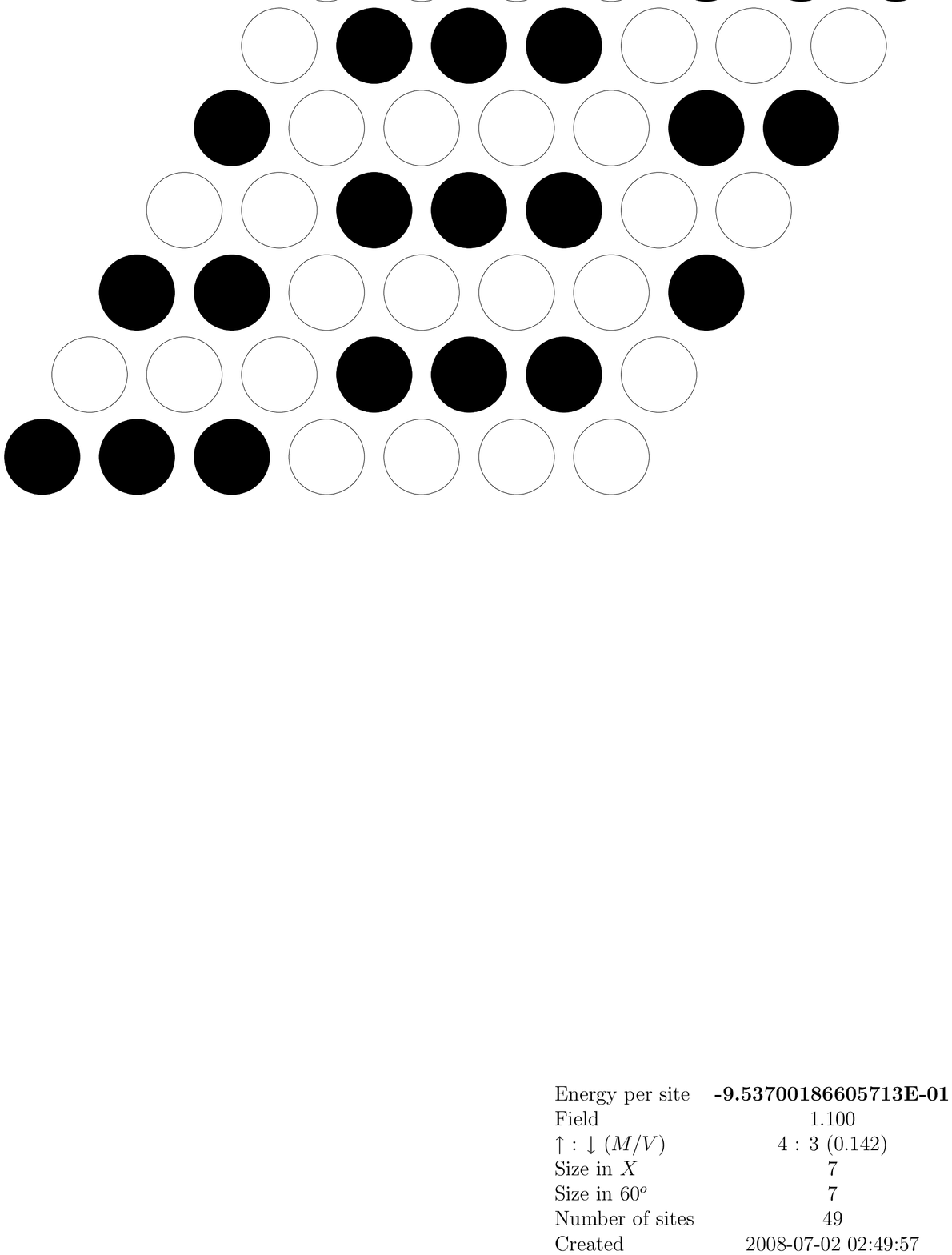}}\hfill \subfigure[\ $h = 1.5$
\label{f:D6z:inf_p5}] {\includegraphics*[bb = 20 500 570 840,
width=\figwidths]{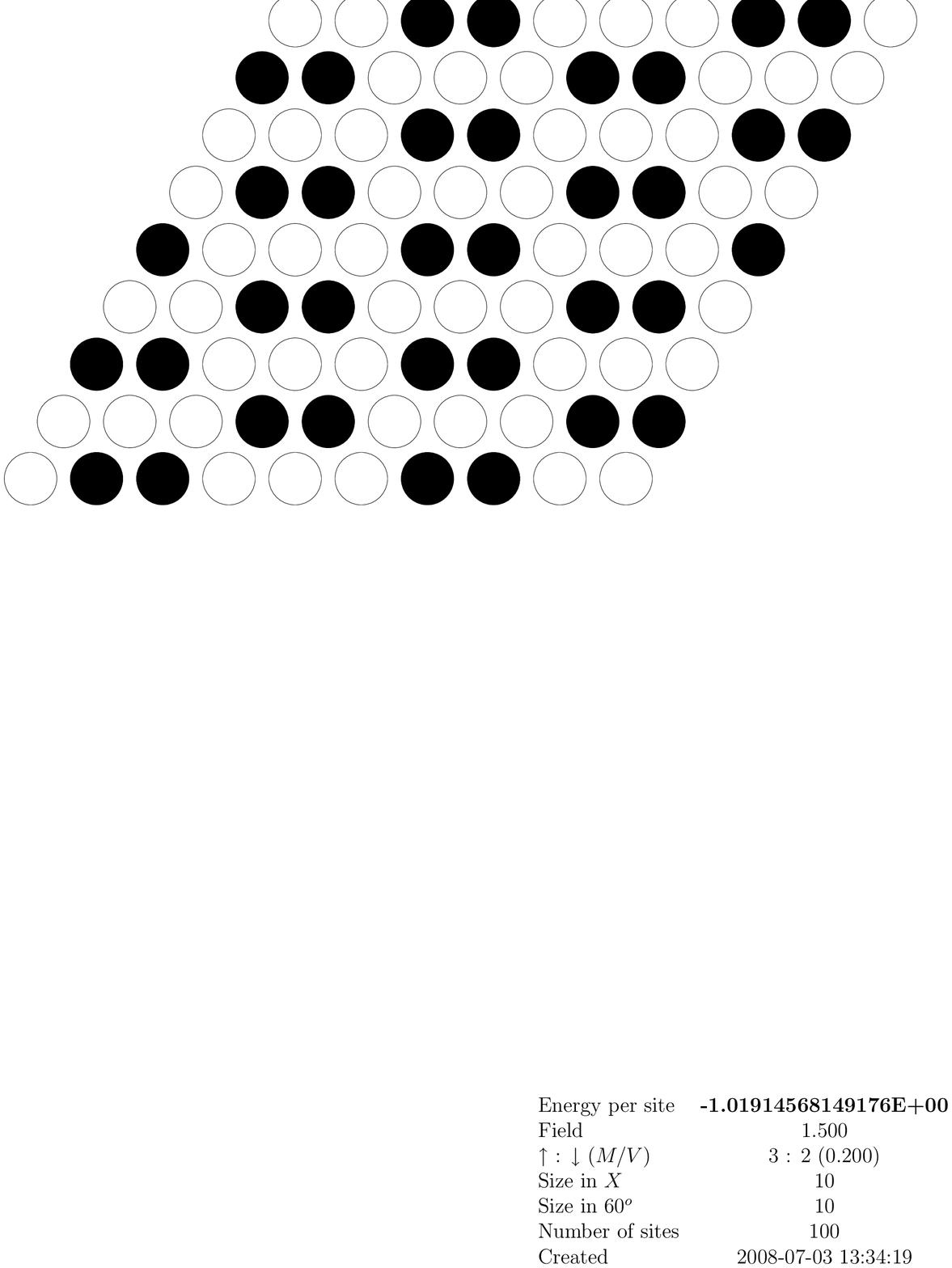}}\hfill \subfigure[\ $h = 2.1$
\label{f:D6z:inf_p8}] {\includegraphics*[bb = 20 500 570 840,
width=\figwidths]{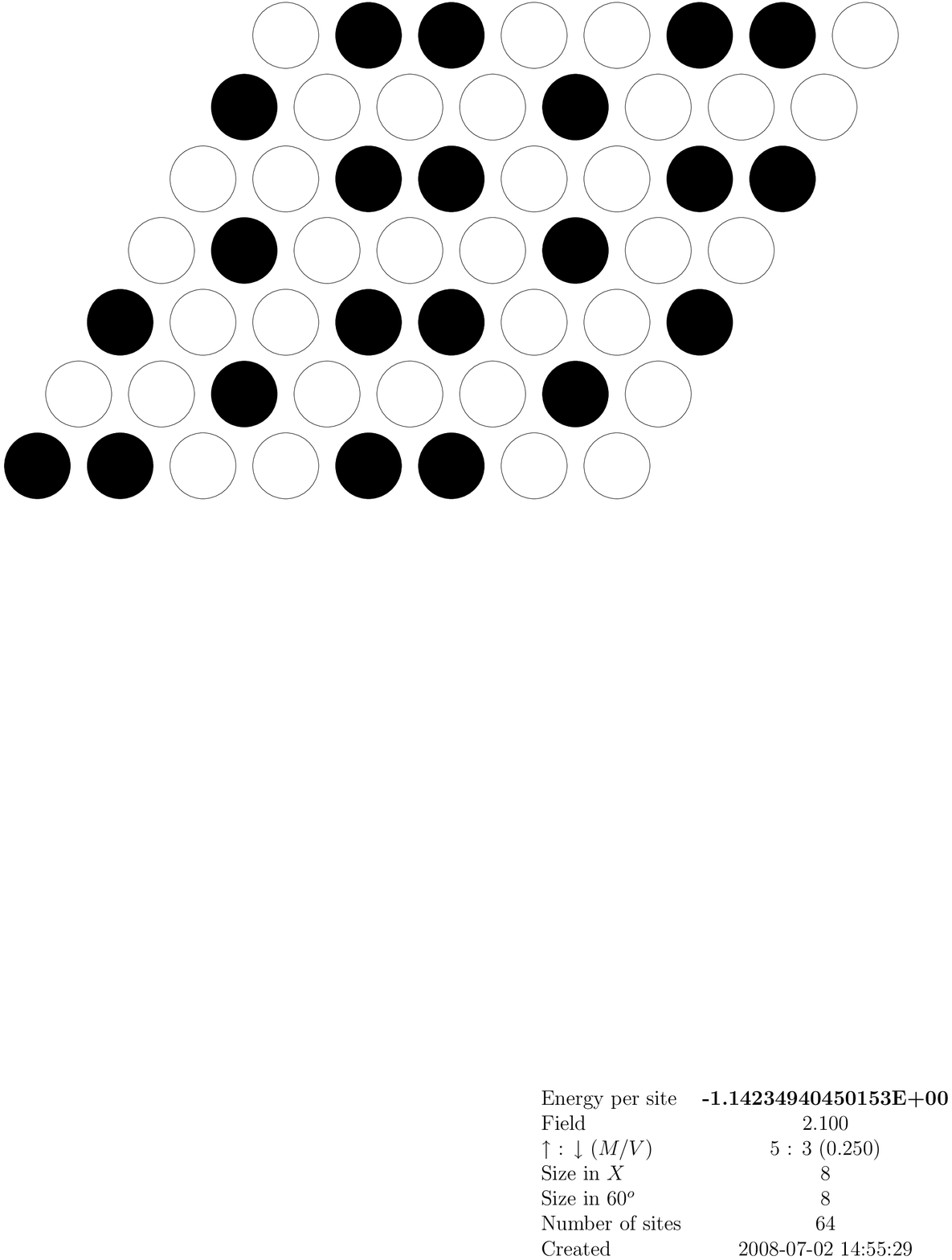}}\hfill \subfigure[\ $h = 2.3$
\label{f:D6z:inf_p11}] {\includegraphics*[bb = 20 500 570 840,
width=\figwidths]{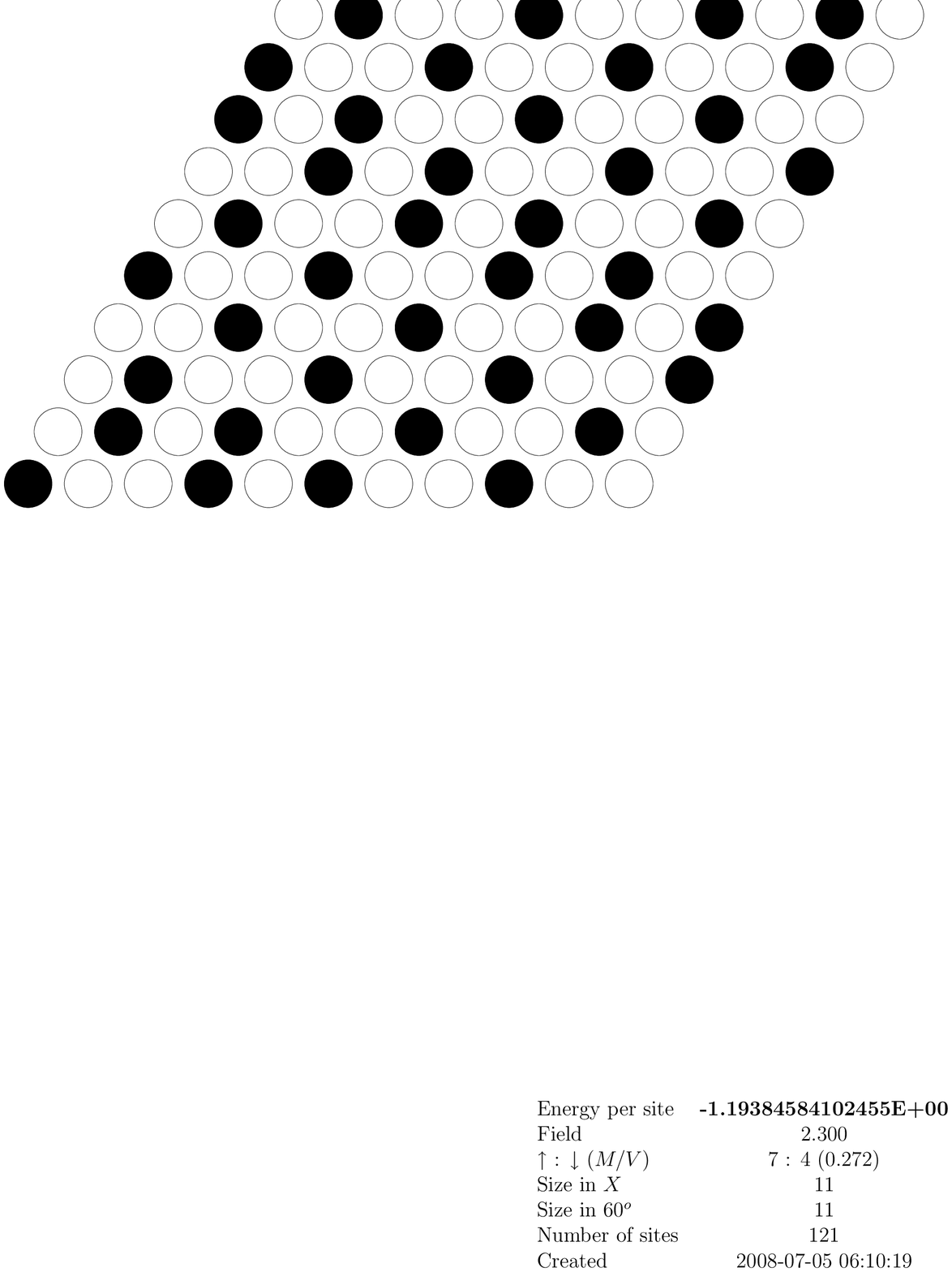}}\hfill \subfigure[\ $h = 4.0$
\label{f:D6z:inf_p3}] {\includegraphics*[bb = 20 500 570 840,
width=\figwidths]{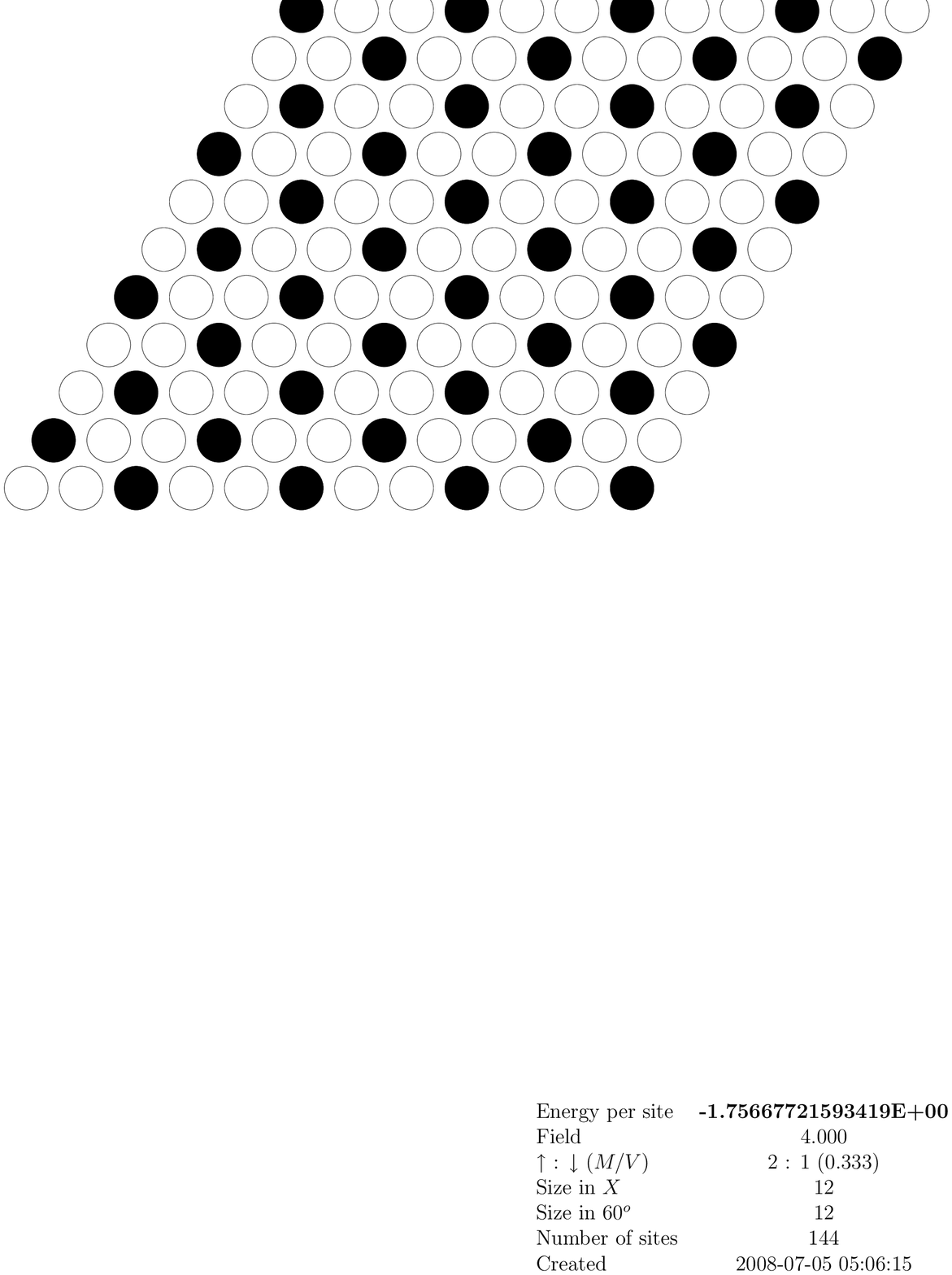}}\hfill \subfigure[\ $h = 6.77$
\label{f:D6z:inf_a3+}] {\includegraphics*[bb = 20 500 570 840,
width=\figwidths]{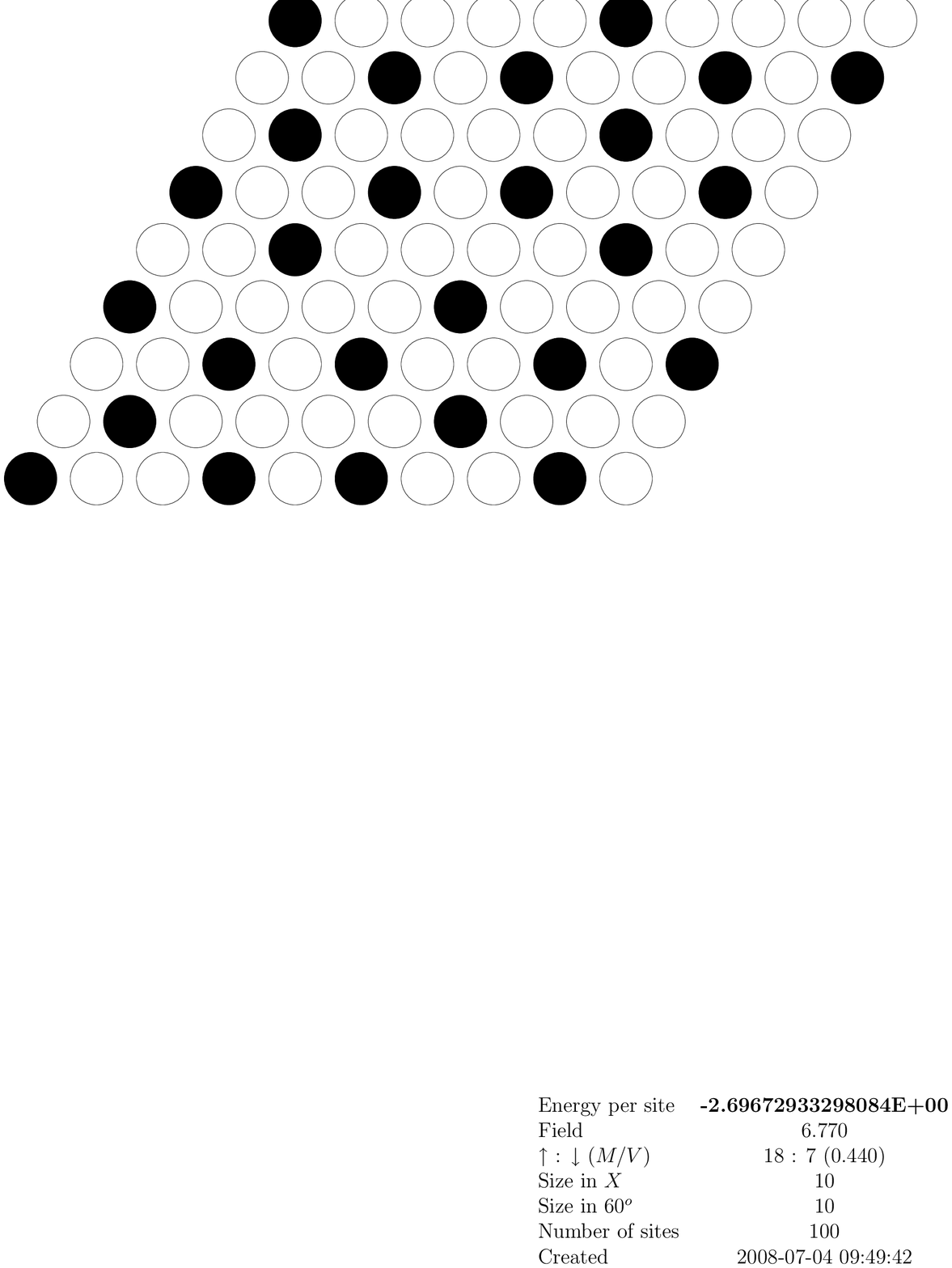}}\hfill \subfigure[\ $h = 7.5$
\label{f:D6z:inf_a4}] {\includegraphics*[bb = 20 500 570 840,
width=\figwidths]{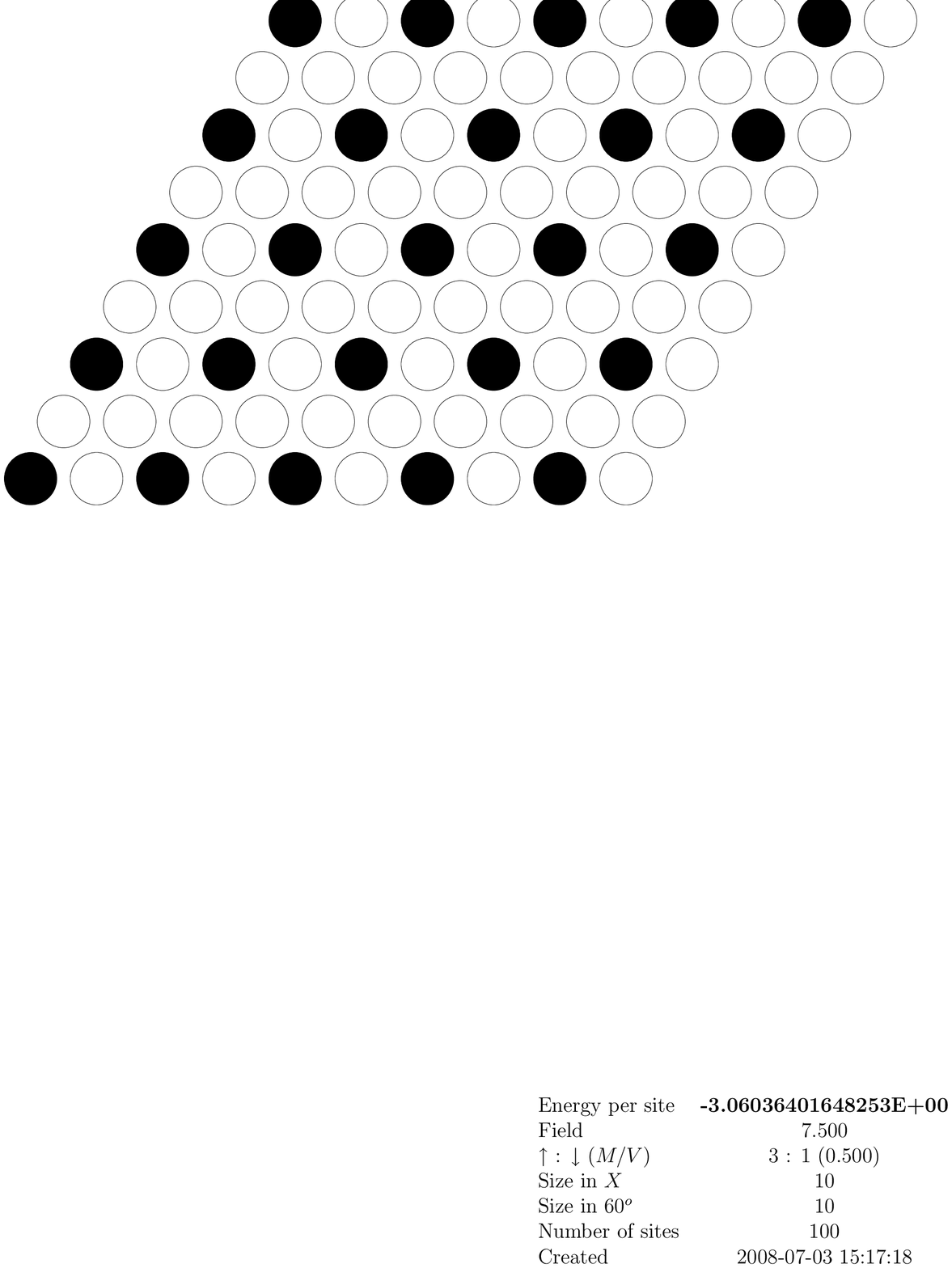}}\hfill \subfigure[\ $h = 9.5$
\label{f:D6z:inf_a7}] {\includegraphics*[bb = 20 500 570 840,
width=\figwidths]{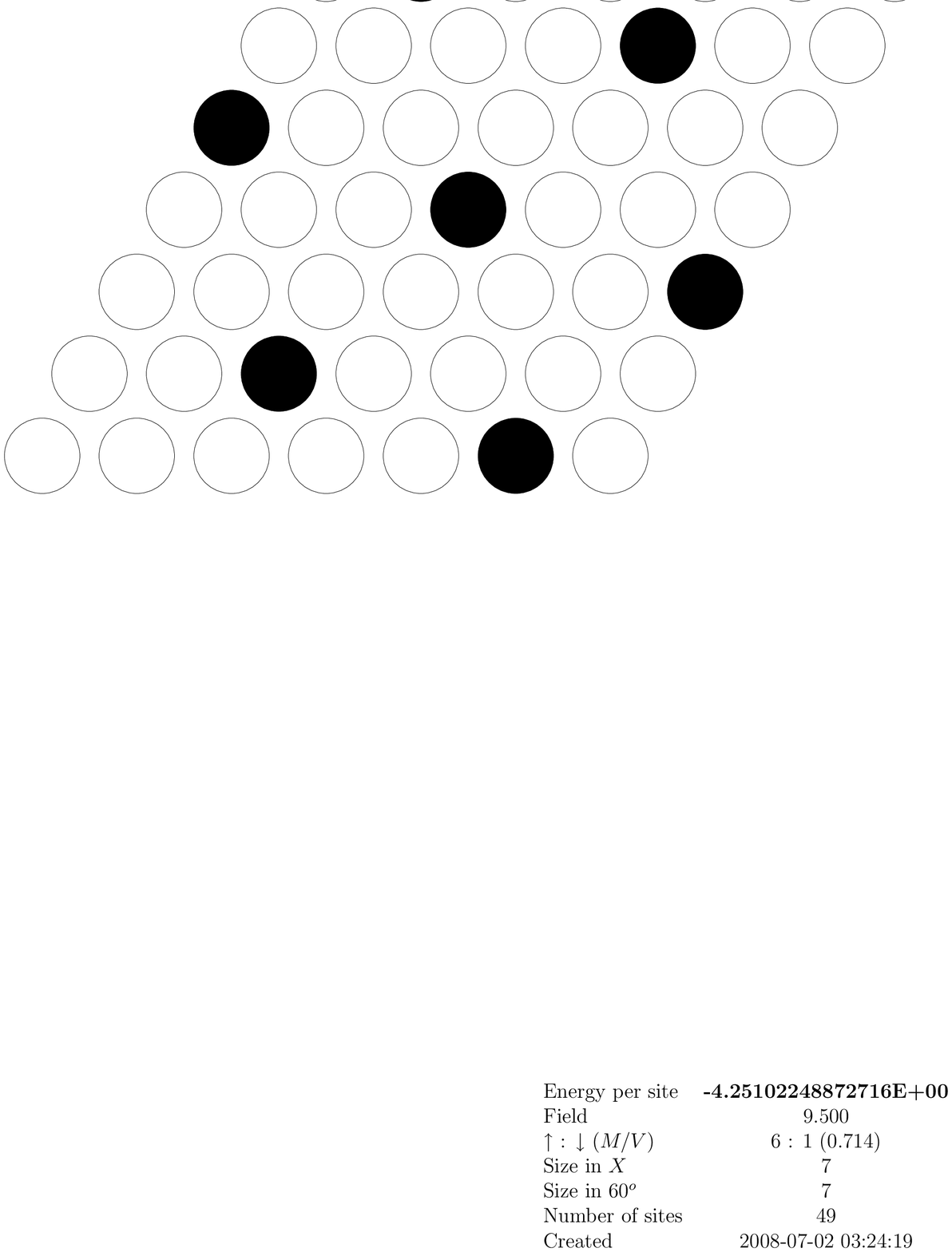}}\hfill \subfigure[\ $h = 10.1$
\label{f:D6z:inf_a9}] {\includegraphics*[bb = 20 500 570 840,
width=\figwidths]{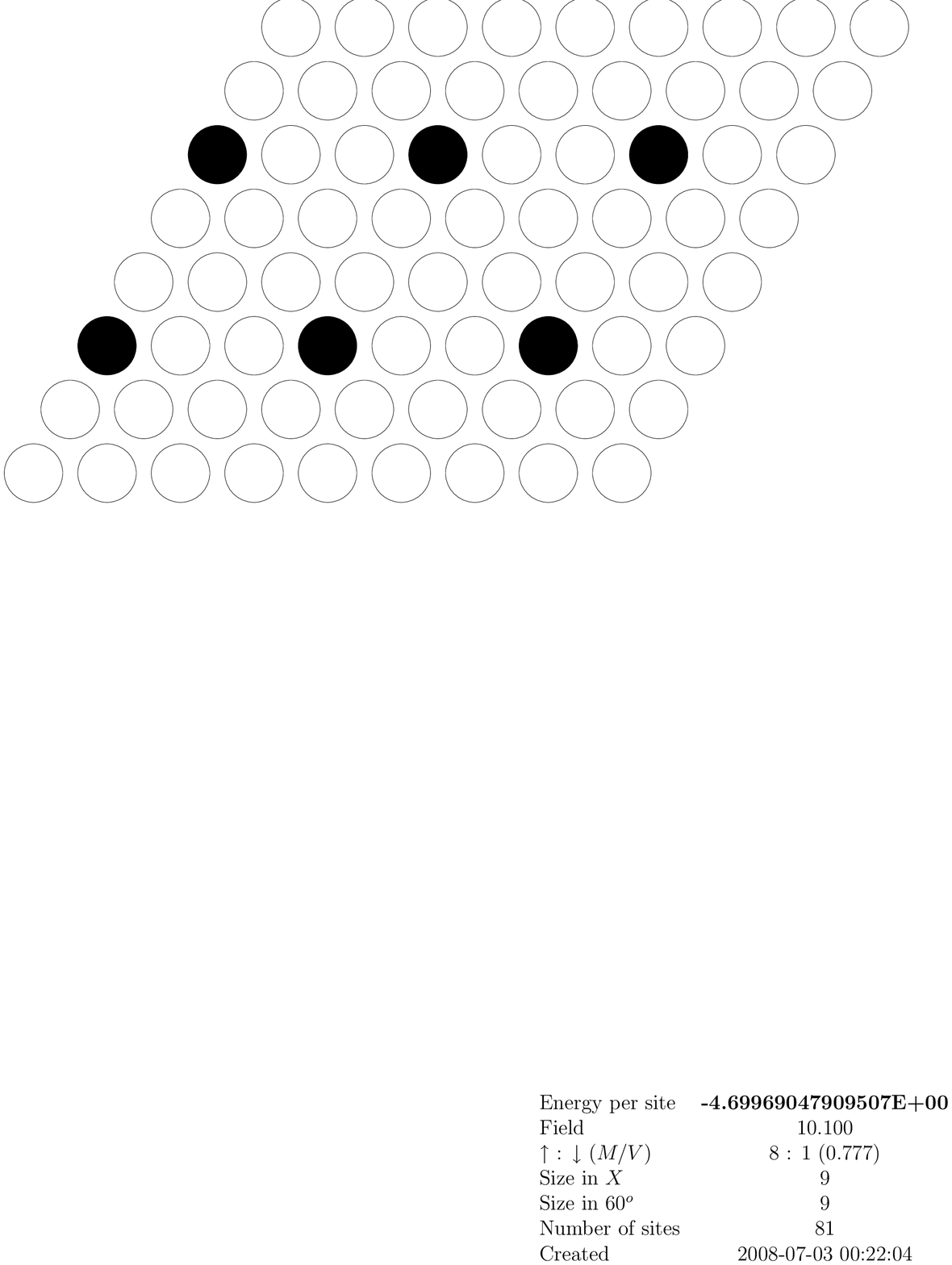}}\hfill \caption{Ground states for
characteristic values of magnetic field found by Monte-Carlo
simulations. \label{f:D6z:inf}}
\end{figure}

Note the specific regions of this dependence, present at different
field intervals; first, the region with small values of $\langle m
\rangle \leq 0.2 m_0$ having rather non-regular dependence of
$\langle m\rangle $ on $h$; second, the regions with constant values
of $\langle m \rangle$ independent on the magnetic field (shelves);
and third, the saturation region. The characteristic magnetic
structures found in these regions are depicted at the
Fig.~\ref{f:D6z:inf}.

Monte-Carlo data are not too clear in the region of small fields
such as 0.7-0.9, and the magnetic structures are far from the simple
AFM structures, see Figs.~\ref{f:D6z:inf_p9} - \ref{f:D6z:inf_p8},
and we will discuss the magnetic structure within this region below.

\subsubsection{Shelves and saturation.}

Within the shelf regions, almost all initial Monte-Carlo
configurations lead to the same magnetic structures, which
correspond to the formation of triangular superlattices for the
minority dots (antiparallel to the magnetic field) with different
lattice spacings. As an example, note the ideal  triangular
superlattices with  $\langle m \rangle = m_0/3$ and with the period
$a_{\rm sl}/a = \sqrt{3}$, see fig.~\ref{f:D6z:inf_p3} present at
the values $2.4 \lesssim h \lesssim 6.4$. For higher fields, the
superlattices with $a_{\rm sl}/a = 2$ (Fig.~\ref{f:D6z:inf_a4}),
$a_{\rm sl}/a = \sqrt{7}$, (Fig.~\ref{f:D6z:inf_a7}) and $a_{\rm
sl}/a = 3$ in Fig.~\ref{f:D6z:inf_a9}) correspond to such shelves.

For the high magnetic fields near the saturation region the
magnetization process is going through the same scenario as for a
square lattice, namely, by a flip of a small amount of magnetic
moments and creating a superlattice of flipped dots of small
density.

\subsubsection{Transition region, topological mechanism. }

In the region of low magnetic fields, as well as in the regions of
magnetic field where the transitions between the superlattices
occur, resettability of Monte-Carlo result is lowering, and the
results becomes unreliable. The observed magnetic structures in
these transition regions are characterized by much lower symmetry
than for the shelf regions.  For example, at the values $0.9
\lesssim h \lesssim 1.5$, where the finite (but small) magnetic
moment $\langle m_0 \rangle$ is formed, the translational symmetry
for the set of flipped dots cannot be attributed to simple
superlattice structure, see Fig.~\ref{f:D6z:inf_p9}. But in this
figure a novel element, the additional zigzag line of the sites,
oriented parallel to one of the translation vector of the lattice,
is clearly seen. The resulting magnetic structure can be interpreted
as an antiferromagnetic domain structure in the system, with the
zigzag line as a domain wall. Such a scenario, magnetization through
the creation of a set of topological linear defects, was described
for a two-sublattice antiferromagnetic state with an interaction of
a few neighboring moments.\cite{IvanovKireev09} For the region of
small fields, the increasing the magnetic field leads to an
increase of the density of the topological linear defects, see
Figs.~\ref{f:D6z:inf_p7} and Fig.~\ref{f:D6z:inf_p5}.

The common ``topological'' scenarios are present for other
transition regions; both below and above the shelf regions with the
magnetic structure of a form of ideal  triangular superlattices of
minority dots. For example, the structure present at $2.0 \lesssim h
\lesssim 2.4$ can be described as a ``compression'' of the domains
of the superlattice of period $a\sqrt{3}$ by the lines of dots with
down magnetic moments, see Figs.~\ref{f:D6z:inf_p8}
and~\ref{f:D6z:inf_p11}, whereas the state at the opposite end of
this shelf can be seen as a ``rarefication'' of the $a\sqrt{3}$
superlattice, see. Fig.~\ref{f:D6z:inf_a3+}. The transition
structures corresponding to the ``higher'' shelves have  $C_6$
symmetry, higher than for low field structures. Note the essential
difference of such topological behavior with what is known for a
square lattice, where the competition of square and triangular
(distorted) superlattices of minority of magnetic dots is
responsible for magnetization processes.

\subsection{Small fields; exhaustive search of the states.}

As has been found by direct Monte-Carlo simulations, in the field
region $0.9 \lesssim h \lesssim 1.5$ the ground state is realized by
a system of parallel AFM stripe domains of the wight depending of
the field, see Figs.~\ref{f:D6z:inf_p9} -- \ref{f:D6z:inf_p5}. The
minimal field for the start of this process corresponds to a low
density of such defects, and to find the critical field one needs to
consider larger and larger system. Namely, to present a stripe of
weight $n$ we need a system of size at least $(2n + 1) \times (2n +
1)$. Below in the Sec.~\ref{sec:analytics} we will find the starting
field for the creation of the set of topological defects by an
analytical calculation.

\begin{figure}
\includegraphics[width=\figwidth]{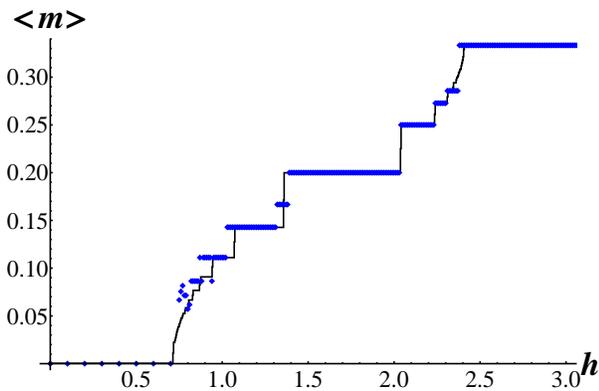}
\caption{Magnetization function at low fields found by exhaustive
search of the states of rhombus-shaped samples (full line) and the
Monte-Carlo data (symbols). \label{f:D6z:magn0_cmp}}
\end{figure}

For refinement the Monte-Carlo data and for clarification of the
magnetic states at the fields of interest, $0 \le h \le 3$, we
perform the direct exhaustive search based on the picture of stripe
AFM domain structures for rhombus-shaped space regions with various
(not necessary equidistant) geometries of domain lines, up to the
size $60 \times 60$. It appears that for all fields the only
equidistant structures corresponds to the minimal configurations,
with linear system of stripes at $h < 1.5$ or triangular
superlattice at $h > 2.0$. Then the only equidistant structures with
the size up to $300 \times 300$ were examined. The magnetization
curve based on these calculations is represented in the
Fig.~\ref{f:D6z:magn0_cmp} and compared with the Monte-Carlo data.

\section{MAGNETIC GROUND STATES: ANALYTIC DESCRIPTION.}\label{sec:analytics}

As has been found in the previous section, for small values of the
magnetic field, the magnetic moment equals zero corresponding to the
simple two-sublattice AFM structure, whereas for high magnetic field
the saturated magnetic structure, or ferromagnetic structure, is
present. It is known that for a non-frustrated square lattice of
Ising magnetic moments the destruction of both types of the magnetic
order is started through creation of the point defect in the state,
the single magnetic dot with the magnetic moment reversed with
respect to the regular structure of a given state.\cite{BishGalkIv}
First, in this section we will check the validity of such scenarios
for ferromagnetic and AFM states of triangular lattice array of
magnetic dots. Then the theoretical description of the novel
topological mechanisms will be obtained.

\subsection{Point defect scenarios. }

In order to determine the values of the magnetic fields which
correspond to such ``point defect instability'' we have calculated
the change in dipolar interaction energy that occurs when the
magnetic moment of a single dot is reversed with respect to the
ferromagnetic and two-sublattice AFM structures. This energy change
is determined by the energy per dot in the initial
states,\cite{BishGalkIv} which can be expressed by simple lattice
sums calculated with high precision. These sums here and below were
calculated with standard program package \emph{Mathematica}.

\subsubsection{Saturation.}
The point defect scenario describes well the instability of the
saturated state. It is easy to see that the change of the energy of
the saturated state with the flip of a single magnetic moment can be
presented as
\begin{equation}\label{D6z:S1_sat}
W_1 = 2m_0 (H -H_{\rm sat})\;,
\end{equation}
where
\begin{equation}\label{D6z:h_sat}
H_{\rm sat}= m_0\sum_{\mathbf{n}_{i} \neq 0}
\frac{1}{|\mathbf{n}_{i}|^{3}} \equiv h_{\rm sat}\frac{m_0}{a^3} \,,
h_{\rm sat} \approx 11.034176 \;.
\end{equation}
The quantity $H_{\rm sat}$ determines the saturation field for a
triangular lattice of Ising magnetic moments. If the magnetic field
$H<H_{\rm sat}$, the value of $W_1$ is negative and the flipping of
a dot becomes favorable. The value $h_{\rm sat}$ is higher than for
a square lattice, $h^{\mathrm{square}}_{\rm sat} = 9.034$, which
just reflects the higher density of the triangular lattice comparing
with the square lattice. Note here that the accounting for only
nearest-neighbor interaction gives a much lower value $h^{NN}_{\rm
sat} = 6$ clearly demonstrating the importance of the long range
character of magnetic dipole interaction for the description of the
ferromagnetic state.

The energy of the simultaneous flip of a pair of magnetic moments
placed at the distance $|\bm{n}|$ can be easily written as
\begin{equation}\label{D6z:S2_sat}
W_2 = 2 W_1 + 4m_0^2 / |\bm{n}|^3 \;,
\end{equation}
that can be interpreted as a repulsion of the flipped dots.

Evidently, for $H<H_{\rm sat}$ some magnetic moments tend to
reverse, and the creation of a finite density of reversed magnetic
moments becomes favorable. The flipped dots should be dispersed as
far from each other as possible, in order to minimize the energy of
their repulsion. Thus one would expect these flipped dots to be
organized in a triangular superlattice, with the lattice spacing
$a_{\rm sl}$. We can regard the resulting system as the
superposition, on the original triangular lattice with magnetic
moment  $+m_0$, of the triangular superlattice of ``double dots'' of
moment  magnetic moment $-2m_0$. From this simple picture, it is
clear that the mean value of the magnetic moment per one dot
$\langle m\rangle$ can be expressed through $a_{\rm sl}$ as
following, $\langle m\rangle =m_0-2m_0(a/a_{\rm sl})^2$.

This picture also allows us to present the energy of dipolar
interaction of the array with a small density superlattice through
the known value of $h_{\rm sat}$, see Ref.~\onlinecite{BishGalkIv}.
First, note that the flipped (double) dots experience the field
$H_{\rm sat}$ from the rest of array. Consequently the interaction
energy of the triangular lattice and superlattice is $2m_0
\varepsilon  H_{\rm sat} $, $\varepsilon = (m_0-\langle m\rangle
)/m_0$, per one dot of the full lattice. Then the contribution to
the energy of each double dot due to its interaction with all other
double dots equals to $(1/2)h_{\rm sat}(2m_0)^2/(a_{\rm sl})^3 $,
per one double dot. This contributes $4W_{FM}(a/a_{\rm sl})^{5/2}$
to the mean dipolar energy per one dot of the overall lattice.
Combining all these contributions and adding the Zeeman energy
$W_H$,  $W_H=-\langle m\rangle H$, we can obtain the approximate
formula for the energy of the superlattice state per one dot in the
form
\begin{equation}\label{D6z:sup_sat}
W=\frac{m_0^2}{a^3}\left[\frac{h_{\rm sat}}{2} - h + \varepsilon
\left(h-h_{\rm sat} \right)+2h_{\rm sat} \left(\frac{\varepsilon
}{2} \right)^{5/2}\right] \,.
\end{equation}
Minimizing the energy \eqref{D6z:sup_sat} over $\langle m \rangle$
(in fact, over $\varepsilon $) yields
\begin{equation}\label{D6z:mh_sat}
\frac{\langle m \rangle }{m_0} = 1 - 2 \left(\frac{2(h_{\rm sat} -
h)}{5h_{\rm sat}} \right)^{2/3} \;.
\end{equation}

This dependence describes very well the numerical  data for the
dependence $\langle m \rangle$ on $H$ near the saturation, which can
be fitted by the dependence common to~\eqref{D6z:mh_sat}, with the
power 0.656 instead of 2/3 $\simeq$ 0.667, and with the coefficient
1.087 instead of $2(2/5)^{2/3}\simeq $ 1.086. Thus the point defect
scenario based on the reversal of a single magnetic moment describes
well the instability of the saturated state for both triangular and
square lattices.

\subsubsection{Low fields.}
In principle, the common calculations can be performed for AFM
state, as well as for any state with the superlattice of flipped
dots of the same symmetry as for underlying dot lattice. This
approach describes well the instability point for the AFM state for
a square lattice of magnetic dots.\cite{BishGalkIv} Let apply it to
our triangular dot lattice. Reversing the magnetic moment of one dot
in the AFM state becomes favorable at $H\geq H_{\mathrm{AFM}}\equiv
2E_{\mathrm{AFM}}/m_0$, where
\begin{multline}
H_{\mathrm{AFM}}=\frac{m_0}{a^3}\sum_{k,l \neq
0}\frac{8}{\left[3\left(2k+1\right)^2
+\left(2l+1\right)^2\right]^{3/2}}-\\ \frac{m_0}{a^3}\sum_{k,l \neq
0} \frac{1}{\left[3k^2+l^2\right]^{3/2}}= 1.8377\frac{m_0}{a^3}
\end{multline}

This value is much higher than the instability field $h \simeq0.7 $
found numerically in the previous section. Thus the reversal of a
single magnetic moment cannot describe the instability of the AFM
state for triangular lattice.

\subsection{Instability of AFM state through creation of topological line.}
\begin{figure}[h]
\begin{center}
\includegraphics[width=0.47\textwidth]{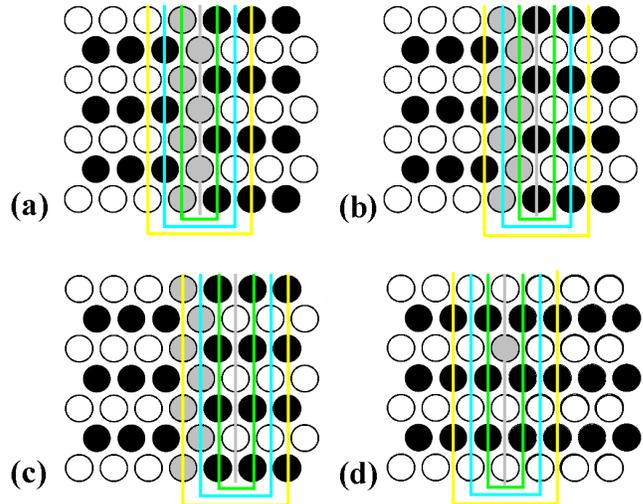}
\end{center}
\caption{(color online). (a) - (c), the structures for topological
defect; (d) the ideal AFM structure; the characteristic lines used
for calculation of the defect energy are shown, see details in the
text. As usually, the open and closed circles denote the particles
with the upward and downward magnetic moments, respectively, but the
dots with upward
magnetic moments within the defect line are mentioned by grey.} %
\label{line_dis} %
\end{figure}

As we found by Monte-Carlo analysis, the AFM state looses its
stability as the field increases because of the creation of
topological defect lines. These lines can be also called domain
walls, because within the description of AFM structure in terms of
the standard antiferromagnetic order parameter (antiferromagnetism
vector $\mathbf{L} = \mathbf{m}_1 - \mathbf{m}_2$, where $
\mathbf{m}_1$ and $\mathbf{m}_2$ are the magnetic moments for
different sublattices) the values of $\mathbf{L}$ have different
signs on both sides of this wall. This  defect line  corresponds to
an additional zigzag line of the particles with magnetic moments,
which are oriented as their neighbors, and should be normal to one
of the elementary translation vectors, see Fig.~\ref{line_dis}.

An importance of defect lines for thermodynamics is a well-known
property of two-dimensional systems with discrete symmetry breaking.
Because of the creation of a finite density of such lines, the long
range order is destroyed at finite temperature, determined by the
energy of the defect, see, for example, the article.\cite{Korshunov}
But for frustrated AFM states the behavior can be very unusual. In
particular, the magnetic order for AFM Ising system with the
nearest-neighbors interaction is absent at any finite temperature $T
> 0$.\cite{NoPT,GekhtObz} Generally, this behavior can be explained
using the defect line picture of the phase transition, with the
vanishing of energy (more exactly, the free energy) of a certain
linear topological defect. For triangular lattice Ising model with
nearest-neighbor interaction it could be linear twin boundary
defects, which separate the different states presented in
Fig.~\ref{f:2-4states}.\cite{Korshunov} Examples of these states
appear for finite systems, see below Sec.~\ref{sec:finite} and
figures wherein. The formation of these defects does not violate the
4–2 relation. Thus their energy is zero in the nearest-neighbor
interaction approximation and should be positive but small for
dipolar coupling.

For our system of mesoscopic magnetic particles with a dipole
interaction, the  effect of a magnetic field, instead of thermal
effects, should be significant. The topological defect line with
non-zero magnetization was recently found.\cite{IvanovKireev09} This
defect coincides with that observed in our numerical simulations,
compare Fig.~\ref{line_dis} and Figs.~\ref{f:D6z:inf_p9} -
\ref{f:D6z:inf_p8} above. The particles directly entering into the
defect line have an unfavorable configuration of the 3–3 type, and
their presence results in energy loss; however, they are adjacent to
the particles (with the downward magnetic moment) having a 5–1
configuration more advantageous than the standard 4–2 configuration.
The numbers of these anomalous states coincide with each other. Thus
the creation of this wall does not result in energy loss in the
system in the nearest-neighbor approximation, and the loss should be
small for particles with dipolar coupling. In contrast to twin
boundary defects considered before,\cite{Korshunov} this domain wall
has a nonzero magnetic moment. As a physical consequences of this
property note that the behavior of the system in a magnetic field is
dictated by the defects with nonzero magnetic moment. In fact, for
such a defect the additional energy gain $m_0H$ per defect particle
appears in the magnetic field $H$. Then the defect energy decreases
as the field increases and becomes zero at $H = H_{\mathrm{DW}}
\equiv E_{\mathrm{DW}}/m_0$. Then for $H \geq H_{\mathrm{DW}}$ the
finite density of such defects will be present in the ground state.
This is exactly the scenario observed in our numerical simulations
at magnetic field at the range 0.7 - 1.5.

In order to find the critical value of the field $H_{\mathrm{DW}}$,
let calculate the energy of the domain wall $E_{\mathrm{DW}}$. It is
convenient to use the symmetry of the state with the defect line and
to divide the full lattice into lines of dots parallel to the defect
line, as it is shown in the Fig.~\ref{line_dis}. It is evident that
all dots within one such line have the same energy, and the energy
of the domain wall (per one dot in the defect line) can be present
as a sum over these lines as following
\begin{equation}
E_{\mathrm{DW}}=-\frac{1}{2}\sum_{n}m_0 \sigma_{n}H_n \label{el_gs}
\end{equation}
where the integer $n$ describes the distance $a_n$ of the given line
from the defect line, $a_n=an/2$, $\sigma_{n}=\pm 1$ gives the sign
of the moment for the $n$-th line, and $H_n$ is the magnetic field
created on the dot in the $n$-th line by other dots in the system.
To find the field $H_n$ it is convenient to group all other dots to
pares of lines, equidistant from the $n$-th line, as it is shown for
$n=0,1,2$ at Fig.~\ref{line_dis}, (a), (b) and (c), respectively.
Let us enumerate these pairs by an integer $k$ so that distance
between $n$-th line and one component of the $k$-th pair equals to
$ak/2$, the pairs with $k=1,2,3$ are present at the
Fig.~\ref{line_dis}. Then the energy of the magnetic state with a
domain wall can be presented by a double sum, over $n>0$ and $k>0$.

It is easy to see that for any finite $n$ the only pairs with
limited $k<n$ contribute to the energy of the state with the domain
wall. For example, for the lines directly entering the defect line
[$n=0$, see Fig.~\ref{line_dis}~(a)] the contributions of two lines
composed any pair cancel each other. For this line, the non-zero
contribution to the energy is given by the dots from the same line,
denote this contribution as $\varepsilon _0$. Then, for the line
with $n=1$, only one pair gives non-zero contribution, see
Fig.~\ref{line_dis}~(b), and the energy can be written as
$\varepsilon _0- 2\varepsilon _1$. Similarly, for $n=2$ the energy
is $\varepsilon _0- 2\varepsilon _1+ 2\varepsilon _2$, see
Fig.~\ref{line_dis}~(c), and so on. Finally, the energy of the state
with domain wall is presented through $\varepsilon _0$ and the
particular finite sums of the positive quantities $\varepsilon _n$,
\begin{eqnarray}\label{Epsn} \nonumber
\varepsilon _{2n+1}&=&\sum_{k=1}^\infty \frac{4}{[(n+1/2)^2 +
3(k-1/2)^2]^{3/2}}\,, \\ \varepsilon _{2n}&=&\sum_{k=1}^\infty
\frac{4}{(n^2 + 3k^2)^{3/2}}+\frac{2}{n^3}\,.
\end{eqnarray}

The energy of the domain wall equals the difference of the energy of
the state with the domain wall and the ground state energy. To find
the ground state energy, it is convenient to use the same
presentation by the parallel lines, see Fig.~\ref{line_dis}~(d), and
to present it by the same sums $\varepsilon _n$. It is clear that
the energy per one dot in any line in the ground state is
proportional to an infinite sum of the form $\varepsilon
_{\mathrm{GS}}=\varepsilon _0+2\sum_{n=1}^{\infty}(-1)^n\varepsilon
_n$. Then the domain wall energy can be found by term-by-term
summation of corresponding contributions of the form $[(\varepsilon
_{0}-\varepsilon _{\mathrm{GS}})+(\varepsilon _{0}-2\varepsilon
_{1}-\varepsilon _{\mathrm{GS}})+...]\equiv
h_{\mathrm{DW}}=2\varepsilon _{1}-4\varepsilon _{2}+6\varepsilon
_{3}+...$. The corresponding infinite series
$h_{\mathrm{DW}}=-2\sum_{n=1}^{\infty}(-1)^n n \varepsilon _n$ are
sign-alternating and converges quite well. Finally, domain wall
energy per one dot $E_{\mathrm{DW}}$ can be presented as follows
\begin{equation}\label{Edw}
E_{\mathrm{DW}}=m_0 H_{\mathrm{DW}}\,,
H_{\mathrm{DW}}=h_{\mathrm{DW}}\frac{m_0}{a^3}\,,
h_{\mathrm{DW}}=0.70858944\,.
\end{equation}

Here we also present the characteristic value of the magnetic field,
and $H_{\mathrm{DW}}=E_{\mathrm{DW}}/m_0$ determining the border of
stability of the simple AFM state; for $H>H_{\mathrm{DW}}$ AFM state
becomes unstable against creation of domain walls. Note the
calculated value \eqref{Edw} is in good agreement with that found by
numerical simulations, but it is much lower than the field of point
defect instability for AFM state, $H_{\mathrm{AFM}}= 1.8377m_0/a^3$.

\subsection{Plateau description}
\begin{figure}[h]
\begin{center}
\includegraphics[width=0.47\textwidth]{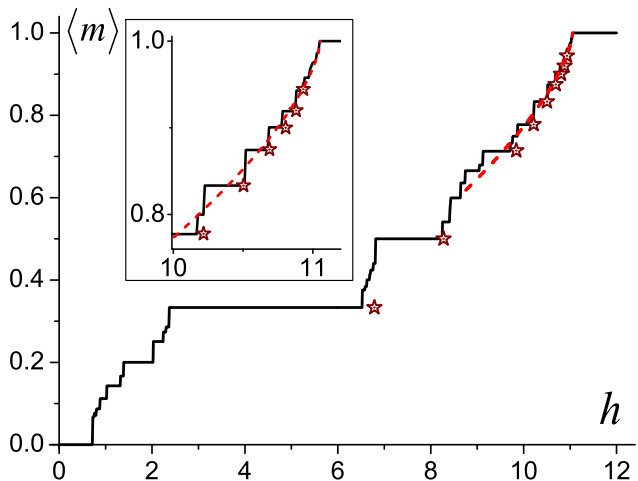}
\end{center}
\caption{(color online) The magnetization function dependence on
external magnetic field. Numerical data are present by solid line,
analytical calculations of
the instability points specified by the symbols, see Eq.~\ref{eqn_step}.} %
\label{plateau} %
\end{figure}

Monte-Carlo simulations result in some peculiarities  (See
Fig.~\ref{plateau}) in the dependence of the magnetization on
applied magnetic field in the form of plateaus, where the value of
the function does not change over a wide range of the argument.

These peculiarities have a simple explanation. The magnetization of
the array increases at small external field due to the formation of
parallel topological defects in the form of domain walls. At some
critical concentration of such walls the resulting state in nothing
but the superlattice of flipped dots which has the triangular
structure that coincides with the array symmetry; See
Fig.~\ref{f:D6z:inf}(f,h,i,j). Such a superlattice transforms into
the structure similar to itself but with the other step (lattice
constant) as the applied field increases. Since the lattice constant
has discrete values  $a_{\rm
sl}/a=\sqrt{3},2,\sqrt{7},3,2\sqrt{3},\sqrt{13},4,\sqrt{19},\sqrt{21},5...$,
then such superstructure has good stability against the alteration
of the external field and magnetization can be changed only
stepwise. One can see that such a structure consists of two
inversely magnetized ferromagnetic states with the lattice constants
$a$ and $a_{\rm sl}$. As stated above the magnetization of a such
state is
\begin{equation}
\langle m\rangle =m_0-2m_0(a/a_{\rm sl})^2.
\end{equation}

The value of the field of the stability loss of such superstructure relative to the transition to other
lattice constant can be easily calculated on the same principle
as the field of the ferromagnetic state stability that was done above.
\begin{equation}\label{eqn_step}
H_{\rm sl}=h_{\rm sat}\frac{m_0}{a^3}\left[1-2\left(\frac{a}{a_{\rm
sl}}\right)^3\right],
\end{equation}
where the multiplier $2$ in the numerator responds to the change of
magnetization of the dot in comparison with the ferromagnetic state
and $h_{\rm sat}=11.034176$ is the field of the transition to the
saturated state. The values of these fields depending on the
superlattice magnetization are represented by symbols (stars) on the
Fig.~\ref{plateau}.

As was found by numerical simulations, the destruction of such
superstructures can go through the creation the lines of topological
defects, see Fig.~\ref{f:D6z:inf}(g). This mechanism assumes that
unit cells of the superlattice repel themselves with an increase of
the field and the line of the dots magnetized inversely to the dots
of the superlattice passes between them. The situation recurs until
a new superlattice formes with a larger lattice constant. Though
this effect occurs in the narrow range of the field it leads to
instability of the superlattice at smaller value of the field than
is predicted in equation (\ref{eqn_step}).

Formation of the triangular superlattice is impossible at small
values of the applied magnetic field $h \leq 2.5$ but some
rectangular superlattices with $C_2$-symmetry are possible, see
Fig.~\ref{f:D6z:inf}(с), which also are the reason for the plateaus
appearance (but not so well pronounced) at the magnetization
function, see Fig.~\ref{f:D6z:magn0_cmp}.

\section{semi-infinite arrays: the role of the boundaries.}\label{semi-infinite}
In the previous theoretical consideration we analyze the idealized
model of the infinite array. Of course, real superstructures are
large, but finite systems.  The border elements (edge surface of an
array) are expected to play a considerable role in the formation of
the properties of the ground state. The existence of any
translational invariance significantly simplifies the problem. For
finite system, there is no translational symmetry. The finiteness of
the system manifests itself in two different ways; first, through
the difference in the coordination numbers for the dot at the border
and in the bulk; and second, by the direct influence of the system
size. In this sections, we will analyze the ``border problem'' only,
assuming that the array is semi-infinite, e.g., it is bordered by
one border line. As well we will discuss the ``edge problem'',
investigating the properties of an edge element, a single dot
located at the crossing of two borders. The systems of finite size
will be considered in the next section, mainly numerically. We
primarily will discuss only the simplest geometry of the system,
supposing that the array's borders are parallel to simplest
translation vectors of the underlying triangular lattice.

\subsection{Analytical description}

Consider the simple two-sublattice AFM state typical for an array at
small enough magnetic fields. For an infinite array, any magnetic
dot is influenced by the magnetic field, generated by other dots of
the array. This field is parallel to the dot's magnetic moment; and
the strength of the field is given by the expression $
h_{\mathrm{AFM}}=1.8377$.

\begin{figure}[h]
\begin{center}
\includegraphics[width=0.47\textwidth]{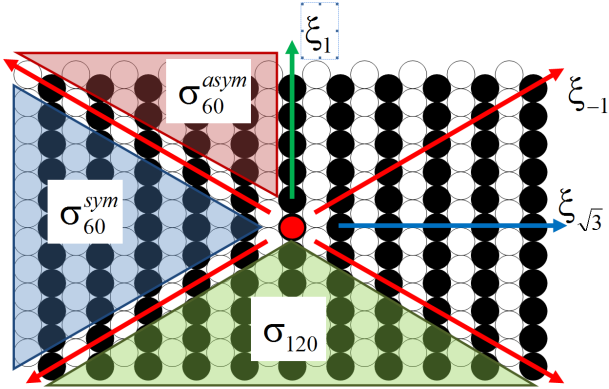}
\end{center}
\caption{(color online) The definition of  auxiliary sums used for
analysis of semiinfinite arrays in AFM state, see the text.} %
\label{bourders} %
\end{figure}

Let consider some different kinds of bordered arrays. It is clear
that the same quantity, the dipole magnetic field at a dot located
near the border, determines the stability of the state. In order to
calculate the field on a given dot, it is convenient to take the
coordinate system with the origin at this dot, see
Fig.~\ref{bourders}. For all cases of interest, namely, for a
magnetic dot located at the edge of the array or at one of its
borders of different orientation, a certain common property is
easily seen. The magnetic field acting at the dot can be expressed
in terms of auxiliary sums, from the single sums  over dots located
at the ray beginning at the coordinate origin (half the coordinate
axis) and a few double sums over dots located in one of the array
sectors that is shown in Fig.~\ref{bourders}. Then for concrete
estimates we will chose the dot at the origin with the ``down''
magnetic moment, so a negative value of the field corresponds to
stability of the configuration and a positive value of the field
 corresponds to the instability of the state.

The single sums are expressed in terms of the Riemann $\zeta(3)$-
function,
\begin{eqnarray}
&& \xi_1=\sum_{n=1}^{\infty}\frac{1}{n^3}=\zeta(3)=1.20206, \nonumber\\
&& \xi_{-1}=\sum_{n=1}^{\infty}\frac{(-1)^n}{n^3}=-3\zeta(3)/4=-0.90154, \\
&&
\xi_{\sqrt{3}}=\sum_{n=1}^{\infty}\frac{1}{\left(3n^2\right)^{3/2}}=\frac{\zeta(3)}{3\sqrt{3}}=0.23134,
\nonumber
\end{eqnarray}
whereas the double sums can be easily determined numerically, for
example,
\begin{equation}
\sigma_{60}^{asym}=\sum_{n=1}^{\infty}\sum_{k=1}^{\infty}\frac{(-1)^n}{\left[n^2
+kn + k^2\right]^{3/2}}=-0.22164
\end{equation}
\begin{equation}
\sigma_{60}^{sym}=-\frac{1}{2}\left[
h_{\mathrm{AFM}}+4\sigma_{60}^{asym}+4\xi_{-1}+2\xi_1
\right]=0.12545.
\end{equation}
Using a simple geometrical consideration, all the double sums of
interest can expressed through the value of $
 h_{\mathrm{AFM}}=1.8377$ and the partial sum
$\sigma_{60}^{asym}$ found above. The knowledge of this particular
sums gives the possibility to determine the stability of any
particular dot of interest.

Let us start with the infinite borders  parallel to the elementary
translation vector. There are two kinds of such a border, with all
magnetic moments parallel and with the alternating magnetic moments,
see  Fig.~\ref{bourders}. The magnetic field on the dot in these
borders, $h_{\uparrow \uparrow}$ and $h_{\uparrow \downarrow}$,
respectively, can be written as
\begin{multline} \label{borders}
h_{\uparrow
\uparrow}=\xi_1- h_{\mathrm{AFM}}/2=0.2832,\\
h_{\uparrow \downarrow}=\xi_{-1}- h_{\mathrm{AFM}}/2=-1.8204.
\end{multline}

Note first that $h_{\uparrow \uparrow}$ is positive, and the
infinite borders  with parallel magnetic moments is unstable in the
absence of the magnetic field, such a fragment can appear at some
finite value of the magnetic field only. Both the aforementioned
properties are in agreement with the numerical analysis of finite
arrays, see the next section. In particular, the border with
parallel magnetic moments never appears for finite arrays; instead,
the complicated multi-domain AFM structure is present. It is worth
noting an essential difference in behavior between triangular and
square lattices. In the latter case, the simple AFM state with
$C_4$-symmetry (chessboard AFM) is the ground state for finite
systems with any shapes of the array including the systems with
acute angles or for  system of circular shape.\cite{GalkinIvMerk}.
In contrast, the ``perfect'' AFM ordering with $C_2$-symmetry for
the triangular lattice is possible only for rhombic shape of the
sample, borders of which are parallel to the fundamental translation
vectors of the lattice. Lines which consist of unidirectional dots
would have sharp bends in all other cases, See Section
\ref{sec:finite}.

The next point of interest is the behavior of the magnetic dot with
the downward magnetic moment at the edge of the array in the AFM
state. For a square lattice, the magnetic moment of such dots
becomes unstable at the magnetic field that is much weaker than the
instability field for the infinite system,
$h^{square}_{edge}=1.563$. Further, the magnetic moment of this dot
is reversed at this value of the field. For a square lattice, such
reversal of the edge dot is the beginning of the destruction of the
AFM state in the finite array. For the triangular lattice, the edges
of finite array are also a ``weak points'' for a single-dot
instability. The values of the instability field for a dot at the
vertex of edges with the angles 60$^{\circ}$ and 120$^{\circ}$,  can
be expressed through the sums $\sigma_{60}^{sym}$ and
$\sigma_{120}$, respectively. The state of the edge dot is then
stable and does not change until the external field increases to the
value. The field $h_{60}$ of the reversal of the single dot at the
vertex of 60-degree edge can be present as
\begin{equation}\label{eq.60}
h_{60}=-(\sigma_{60}^{sym}+2\xi_{-1})=1.67764,
\end{equation}
whereas for 120-degree edge the instability field $h_{120}$ has the
lower value,
\begin{equation} \label{eq.120}
h_{120}=-(2\sigma_{60}^{asym}+\xi_1+2\xi_{-1})=1.0443
\end{equation}
Thus we found that the more sharp edge appears to be more stable
against the action of the magnetic field (note that for square
lattice the situation is opposite\cite{GalkinIvMerk}). This result
seems to be a contra-intuitive, but it is the reflection of the
frustration for AFM state at the triangular lattice.

The above field for single-dot edge and border instability are
weaker than the instability field for the border of infinite system
$h_{\uparrow \downarrow}=1.82039$ but their values are higher than
the value $h_{DW}=0.7085$, where the creation of the domain wall
becomes favorable. Thus, the scenario with the creation of such
domain walls is preferable in a thermodynamic limit. On the other
hand, the creation of a topological domain wall require the
perturbation with a reversal of an essential number of dots
(formally, proportional to the system area). Thus such a process
needs the overcoming of a very high (formally, infinite) potential
barrier, and it is hard to realize it for a real experiment.

\begin{figure}[h]
\begin{center}
\includegraphics[width=0.47\textwidth]{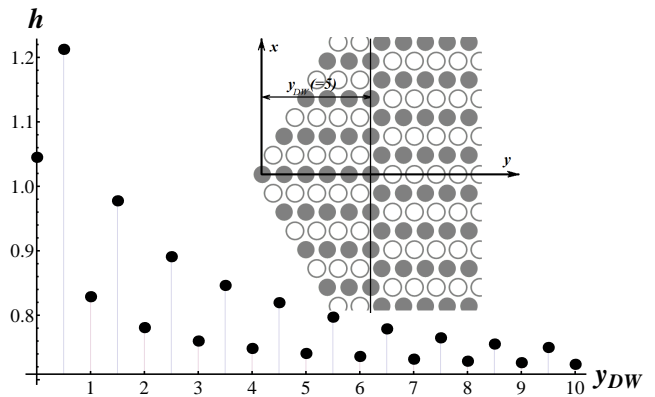}
\end{center}
\caption{Dependence of the field of line formation on the distance to the corner of the array.} %
\label{cornerH} %
\end{figure}

To find an alternative scenario, let us consider the domain wall
which pass nearby the 120-degree corner. Thereto we fix the corner
site to the origin of coordinates and build $y$-axis on the bisector
of the angle, see insert on the Fig.~\ref{cornerH}. Then the domain
wall which lays at the distance $y_{DW}$ results in the
magnetization flip-over, in comparison with the ground state, on all
sites left from the wall. Single dot on the vertex angle is
equivalent to the domain wall with the coordinate $y_{DW}=0$. We
calculated the necessary field of such a domain wall appearance for
different $y_{DW}$, see Fig.~\ref{cornerH}. Calculations were
carried out by the direct summation of the array energy, which
consists of 400 lines in the $y$ direction.

As one can see from the the data on Fig.~\ref{cornerH}) the
necessary field is higher for the domain wall near the corner of the
array than for the deep-laid wall. Also, we can see from the figure
that the field of the domain wall generation depends on the relative
directions of the applied magnetic field and the magnetization of
the vertex site (taken, as before, negative on the figure). Thus
only domain walls with integer values of $y_{DW}$ are of interest
for the problem. With an increase of $y_{DW}$ the field that is
required for the wall formation decreases converging exponentially
to the field of creation of infinite wall, that is similar to the
presence of the potential barrier on the surfaces of solid.
Therefore one can assert that the smaller the difference between
applied field and $h_{DW}=0.7085$ the farther from the corner domain
wall appears.

It is necessary to take into account for large arrays that domain
wall at the sufficient distance $|y|$ from the site produces the
magnetic field at this site proportional to
\begin{equation}
\int_{0}^{\infty}\frac{dx}{\left(x^2+y^2\right)^{3/2}}
=\frac{2}{y^2},
\end{equation}
notably converges to zero when $|y|\rightarrow \infty$. Therefore,
the formation of the another deep lying domain wall is more
energetically favorable than the wall near the corner for the small
concentration of topological defects. And consequently the
magnetization of the large arrays increases firstly in the volume,
where equilibrium density of parallel topological defects would be
observed; and only in the case of external field considerably
exceeding $h_{DW}=0.7085$ domain walls go up to the corner. This
case is opposite the magnetization process for a square lattice
which is initialized on the borders. If we go out of the
thermodynamical limit and look at the slow dynamics of the system
then we could expect that for the strong external magnetic field
(close to $h_{120}$) the topological defects initially appears close
to the corner and then drifts to the volume due to the aforesaid
reduction of the potential barrier.

\section{ground state for finite arrays}\label{sec:finite}

An analysis of finite arrays was accomplished numerically, with
usage of the same Monte-Carlo procedure as described above for
infinite systems. The different shapes, either fully consistent with
the geometry of triangular lattice (bordered by  dense lines of
magnetic dots only, like rhombus, triangle, hexagon) or less
consistent with underlying lattice (rectangle, circle) shows a big
variety of the magnetic structures.

\subsection{Zero magnetic field}
As has been found in the previous section, the simple two-sublattice
AFM survives near the border of semi-infinite system, with one
essential exception: the border of the type of a dense line with
parallel magnetic moments is unstable at zero magnetic field. This
instability is crucial for an understanding of the formation of the
ground state for finite samples bordered by dense lines only
(rhombus, triangle and hexagon).
\subsubsection{rhombus}
\begin{figure}[h]
\begin{center}
\includegraphics*[bb = 20 482 570 830, width=0.2\textwidth]{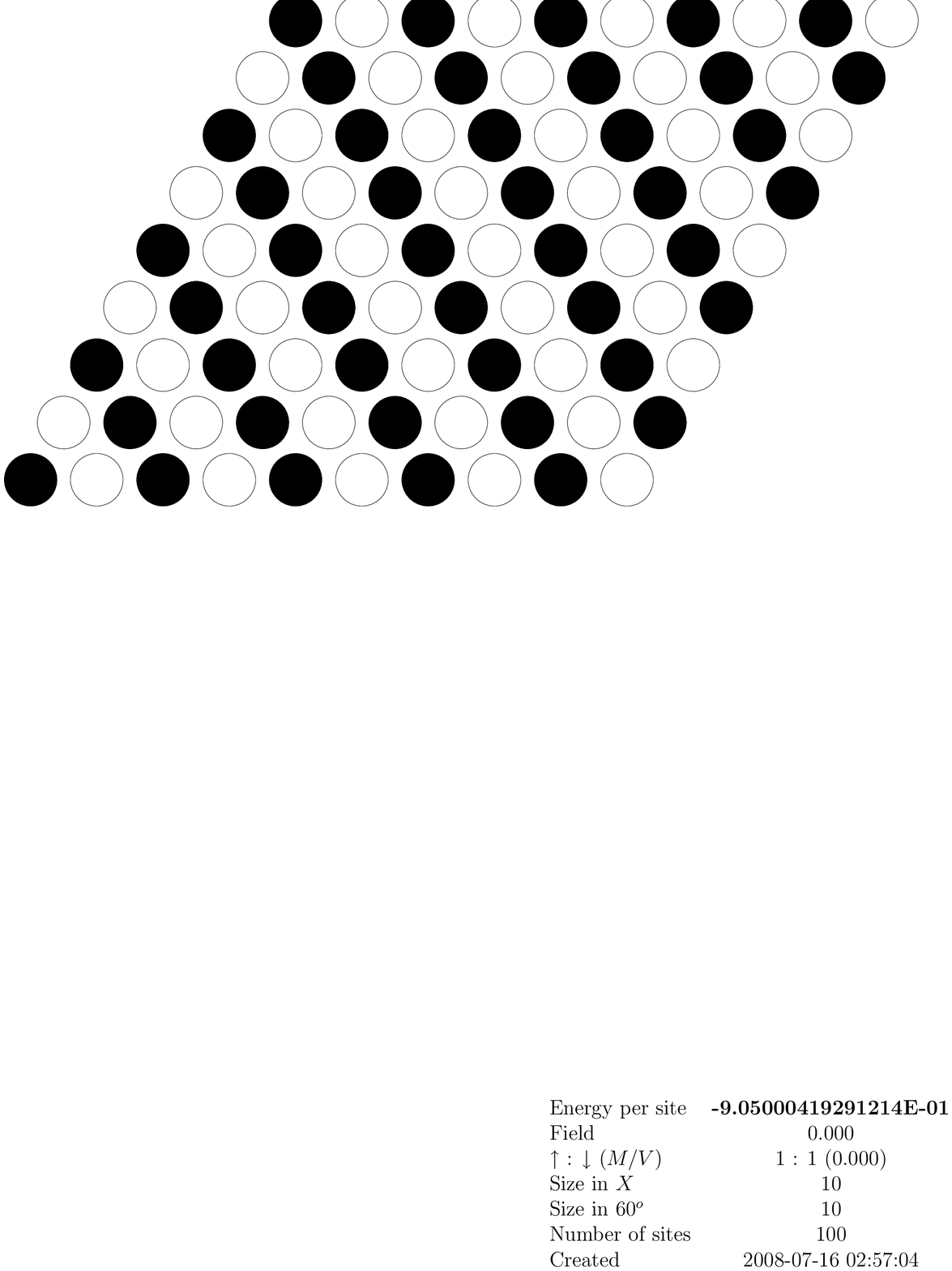}
\end{center}
\caption{Minimal configuration for a rhombus-shaped finite system at zero magnetic field.} %
\label{f:D6z:rhomb10} %
\end{figure}

The rhombus is the only simple geometrical form of finite array with
the simple AFM structure which can be bordered by dense lines for
particles with alternating magnetic moments. Here the three
possibilities of orientation of rhombus with respect to the lattice
corresponds with three alternating AFM structures present at
Fig.~\ref{f:2-4states}. For this reason, at zero magnetic field only
the rhombus could provides ideal AFM ordering. Numerical analysis
confirm this suggestion demonstrating simple two-sublattice
structure, with lines from one sublattice inclined by the angle
60$^\circ$ to the border lines, for rhombuses with different sizes
and shapes, see an example on the Fig.~\ref{f:D6z:rhomb10}.

\subsubsection{triangle and hexagon}
\begin{figure}
\includegraphics*[bb = 20 350 580 830, width=0.2\textwidth]{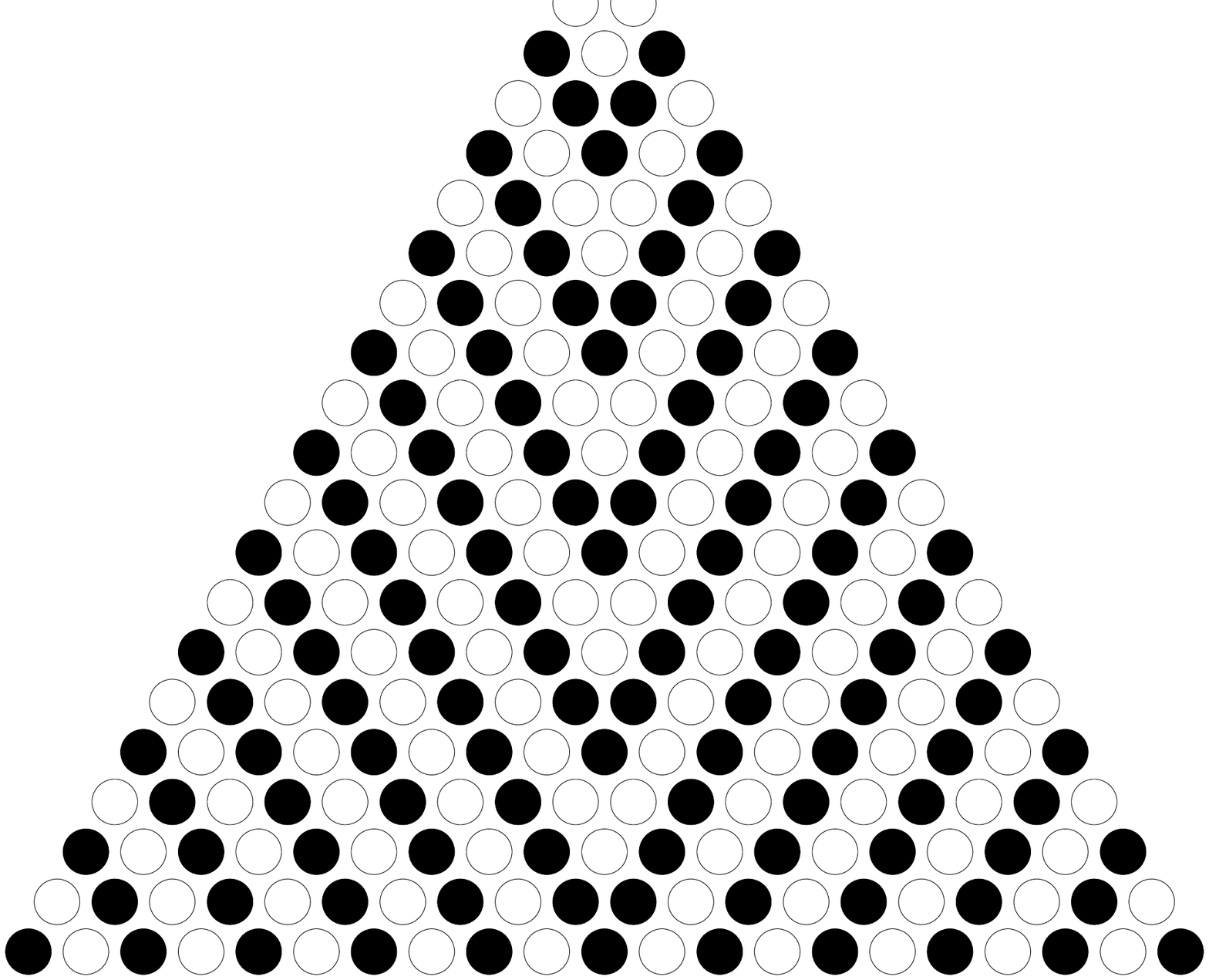}
\includegraphics*[bb = 20 330 570 800, width=0.2\textwidth]{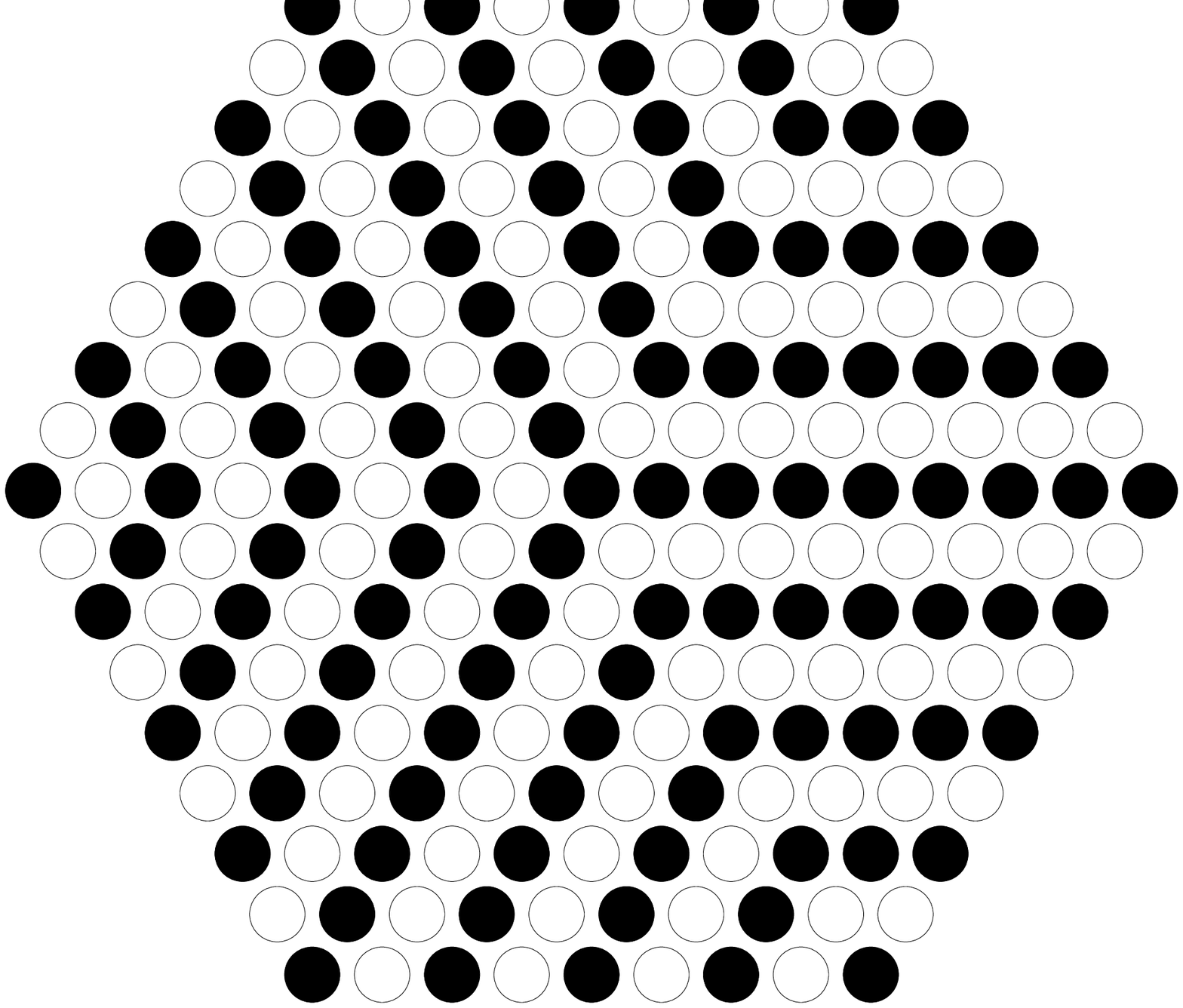}
\caption{minimal configurations for the triangle with the side 21
(left) and for the hexagon with the  side 8 (right).
\label{f:D6z:tri21+hex8}}
\end{figure}

The samples shaped as a right triangle or a hexagon with ideal AFM
structure cannot satisfy the above condition, an absence of the
border line from one sublattice. This leads to appearance of some
kind of domain structures in the ground state, see
Fig.~\ref{f:D6z:tri21+hex8}.

For triangular sample, the domain structure with two AFM domains
turned at the angle of 60$^\circ$, is present in the ground state.
For the hexagon, the minimal configuration consists on three
domains, where the sublattice lines are inclined on 120$^\circ$. The
three domain walls in this structure are going out from the corners
and coming together at the center of the sample in such way that the
structure has the three-fold symmetry axis, see right part of the
Fig.~\ref{f:D6z:tri21+hex8}.

It is worth noting here that the corresponding domain walls have
zero energy in the nearest-neighbor approximation, but the energy
losses caused by the walls are small for long-ranged dipole
interaction as well. For these samples most of the <<bulk>> sites
have the optimal <<2-4>> neighborhood, with the only one exception:
The <<2-4>> rule is broken for the central site of hexagonal sample.
In the nearest-neighbor approximation, the effective field on the
central site equals to zero, and the central magnetic moment is very
sensitive to the application of the magnetic field. For large
hexagonal sample with real magnetic dipole interaction, the field of
the turn of this moment is also small,
\begin{equation}
H_{hex}=-(6\sigma_{60}^{asym}+3\xi_1+3\xi_{-1})\frac{m_0}{a^3}=0.428266\frac{m_0}{a^3}
\,,
\end{equation}
and the magnetization reversal is starting by the turn of the
central magnetic moment. But detailed analysis shows that the field
for the turn of the moments neighboring the center is growing fast
with the distance to the center, and the magnetization reversal is
following the same scenario as for infinite or semi-infinite
systems, see the next subsection.

\subsubsection{rectangle}

\begin{figure}[t]
 \subfigure[\ . \label{f:D6z:rect_x}]
{\includegraphics*[bb = 20 590 570 830,  angle=90,
width=0.1\textwidth]{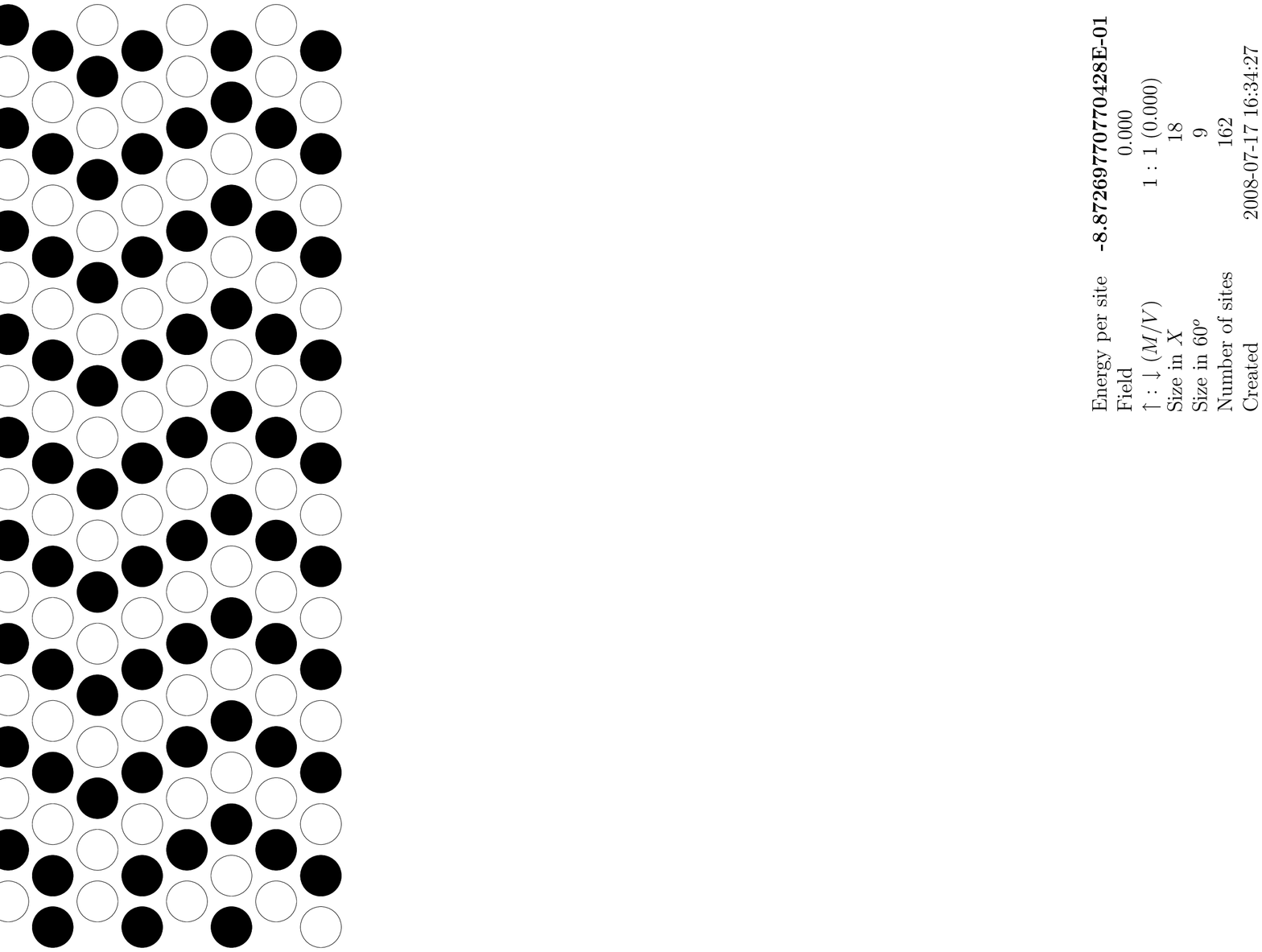}}\hfill \subfigure[\ .
\label{f:D6z:rect_xy}] {\includegraphics*[bb = 20 370 570 830,
width=0.15\textwidth]{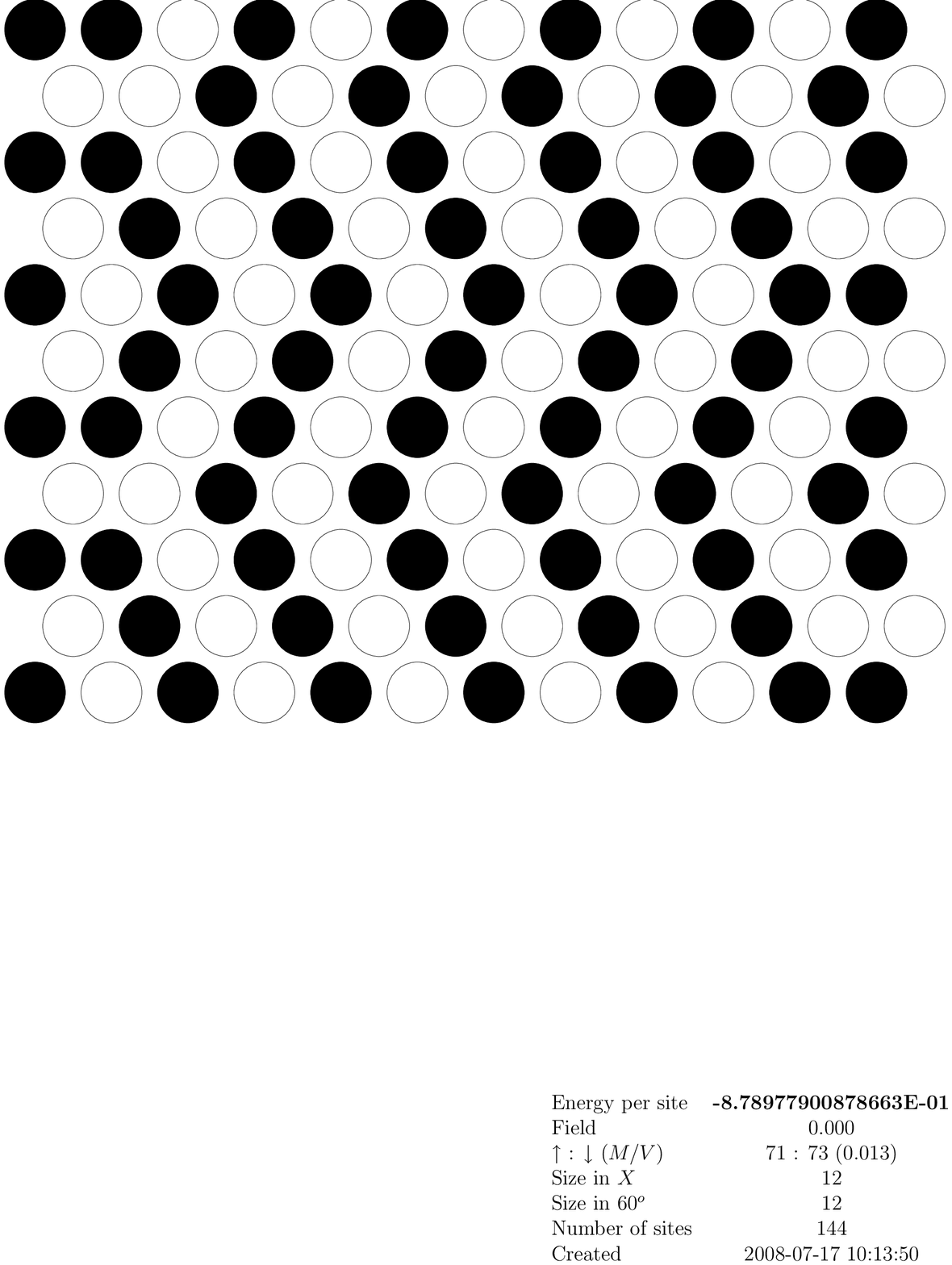}}\hfill \subfigure[\
\label{f:D6z:rect_y}] {\includegraphics*[bb = 20 80 470
830,
width=0.15\textwidth]{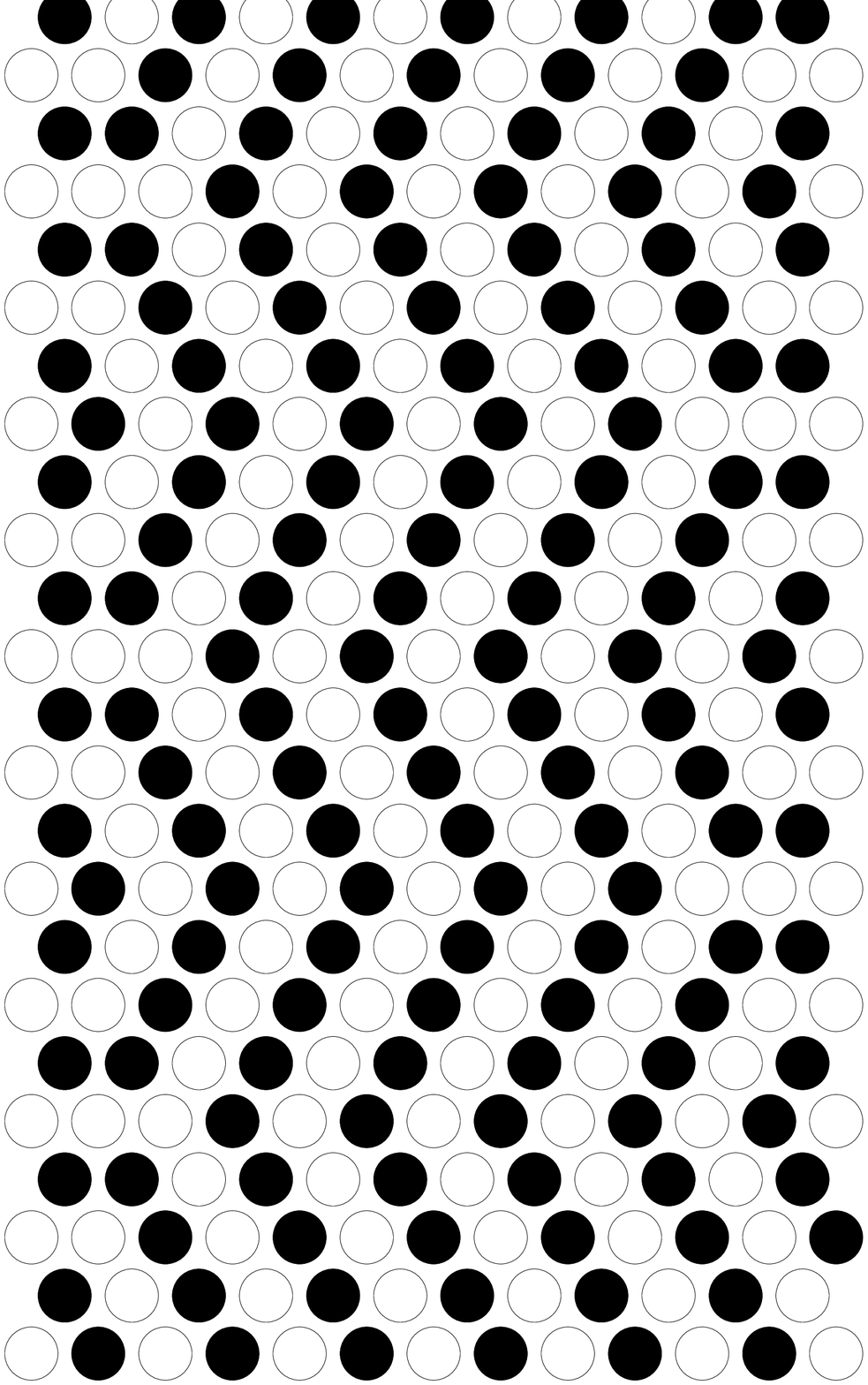}}\hfill \caption{Minimal
configurations for rectangular samples of different shapes; (a)
---  rectangle, elongated along the elementary translation  vector;
(b) --- square sample, (c) ---  rectangle, elongated perpendicularly
to the elementary translation vector. \label{f:D6z:rect}}
\end{figure}

For trivial geometrical conditions, the rectangular sample can have
only one pair of sides (two parallel sides) parallel to elementary
translation vector; these sides have standard AFM structure with
interlacing of up and down magnetic moments, see
Fig.~\ref{f:D6z:rect}. Two other pair of borders of rectangular are
parallel lines of sites with relatively low density within the line
(the distance between the sites equals to $a\sqrt{3}\simeq 1.73a$),
but with a small enough distance $(a/2)\sqrt{3}<a$ with the next
equivalent lines. The preferable type of ordering for two such lines
corresponds to the parallel alignment of the magnetic moments within
the line with the antiparallel orientation for neighboring lines
(<<2-0>> type surface state) whereas the ideal AFM structure leads
to much less favorable <<1-1>> type surface state. For this reason,
one can expect the deformation of the ideal AFM structure in a
rectangular sample. The numerical analysis confirms this simple
speculation based on nearest-neighbor approximation, see
Fig.~\ref{f:D6z:rect}.

The ground state of  rectangular sample contains zigzag deformation
of the characteristic lines of the ideal AFM structure such that the
ideal AFM sign-interlacing distribution is kept at the dense border,
with the specific structure of finite half-hexagons on the less
dense borders. This structure provides the aforementioned picture of
saturated lines parallel to less dense borders with <<2-0>> type
surface state. The zigzag lines are always parallel to the less
dense borders, whereas the hexagon surface structure depends on how
the long side of rectangular sample is oriented respectively to the
elementary translation vector, see Fig.~\ref{f:D6z:rect_x} and
Fig.~\ref{f:D6z:rect_y} for elongated rectangles and
Fig.~\ref{f:D6z:rect_xy} for a square sample.

\subsubsection{circle}

\begin{figure}
\includegraphics*[bb = 20 310 560 830, width=0.2\textwidth]{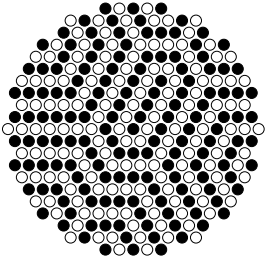}
\caption{Ground state magnetic configuration for the circular sample
of the radius 9 found by Monte-Carlo simulations.
\label{f:D6z:circ9}}
\end{figure}

The circular samples are considered here as an example of systems
with non-small surface roughnesses. The numerical analysis
demonstrates the essential deformations of the ideal AFM structure
for such  samples, see Fig.~\ref{f:D6z:circ9}.  Quite non-regular
border structure having sites with 3 or 4 nearest neighbors, leads
to big variety of the local surface configurations with different
numbers of parallel and antiparallel nearest magnetic moments, like
<<2-1>>, <<3-0>> and <<3-1>> local states. But note that the local
AFM structure, in particular, <<4-2>> condition for the local
states, is kept for most bulk sites.

\subsection{Magnetization processes for finite samples.}

\begin{figure}
\includegraphics[width=\figwidth]{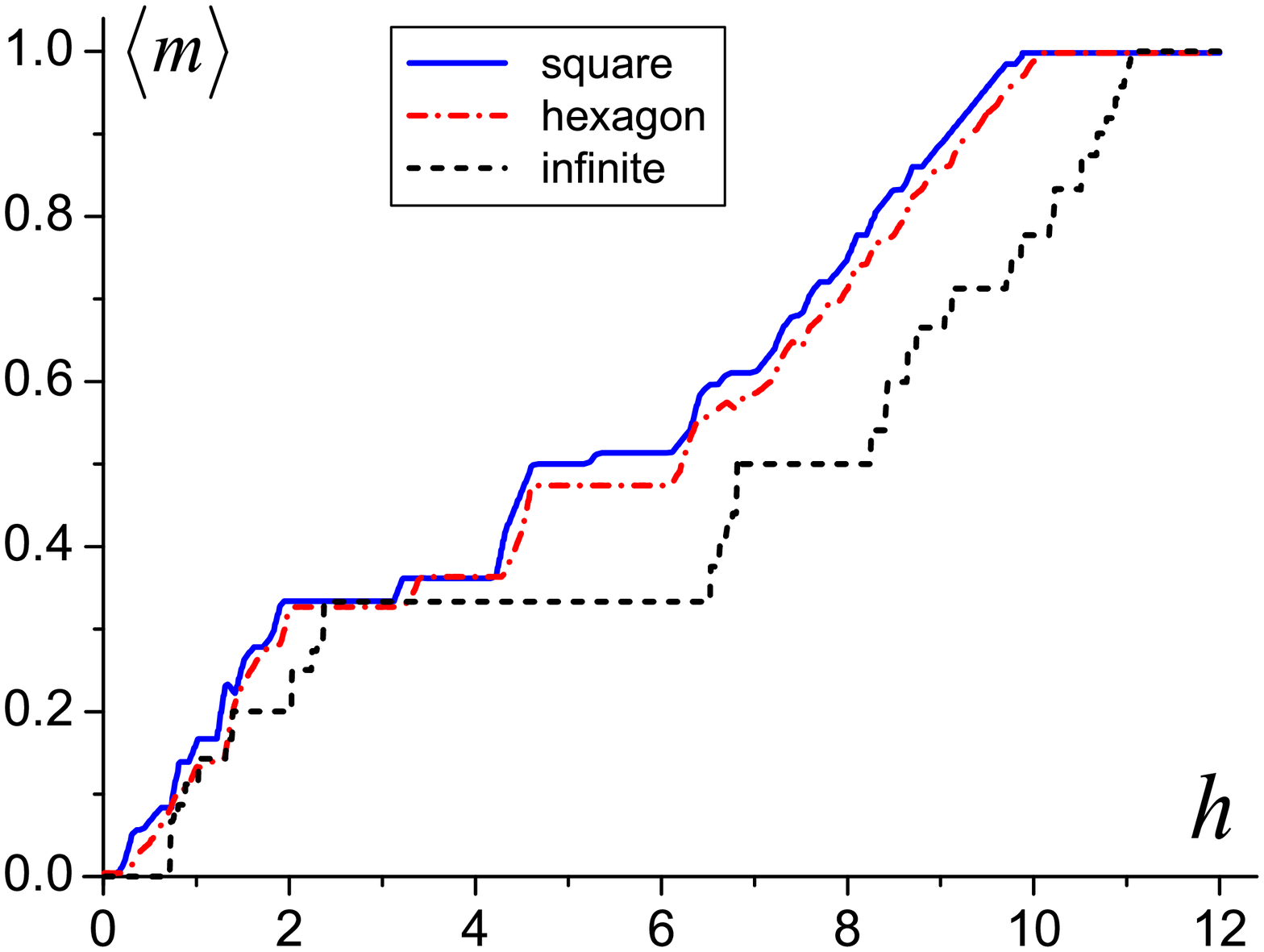}
\caption{(color online) Magnetization function for the finite
systems; hexagon with the side length 8 (215 sites) and square $12
\times 12$ (144 sites), in comparison with that for infinite system.
\label{f:D6z:magn_fin}}
\end{figure}

The magnetization functions for finite systems: hexagon with the
side length 8 (215 sites) and square $12 \times 12$ (144 sites) are
presented on the Fig.~\ref{f:D6z:magn_fin}. The result for infinite
system is cited for comparison.

\begin{figure}
\subfigure[\ $h = 1.0$ \label{f:D6z:hex0}] {\includegraphics*[bb =
30 340 565 840, width=\figwidths]{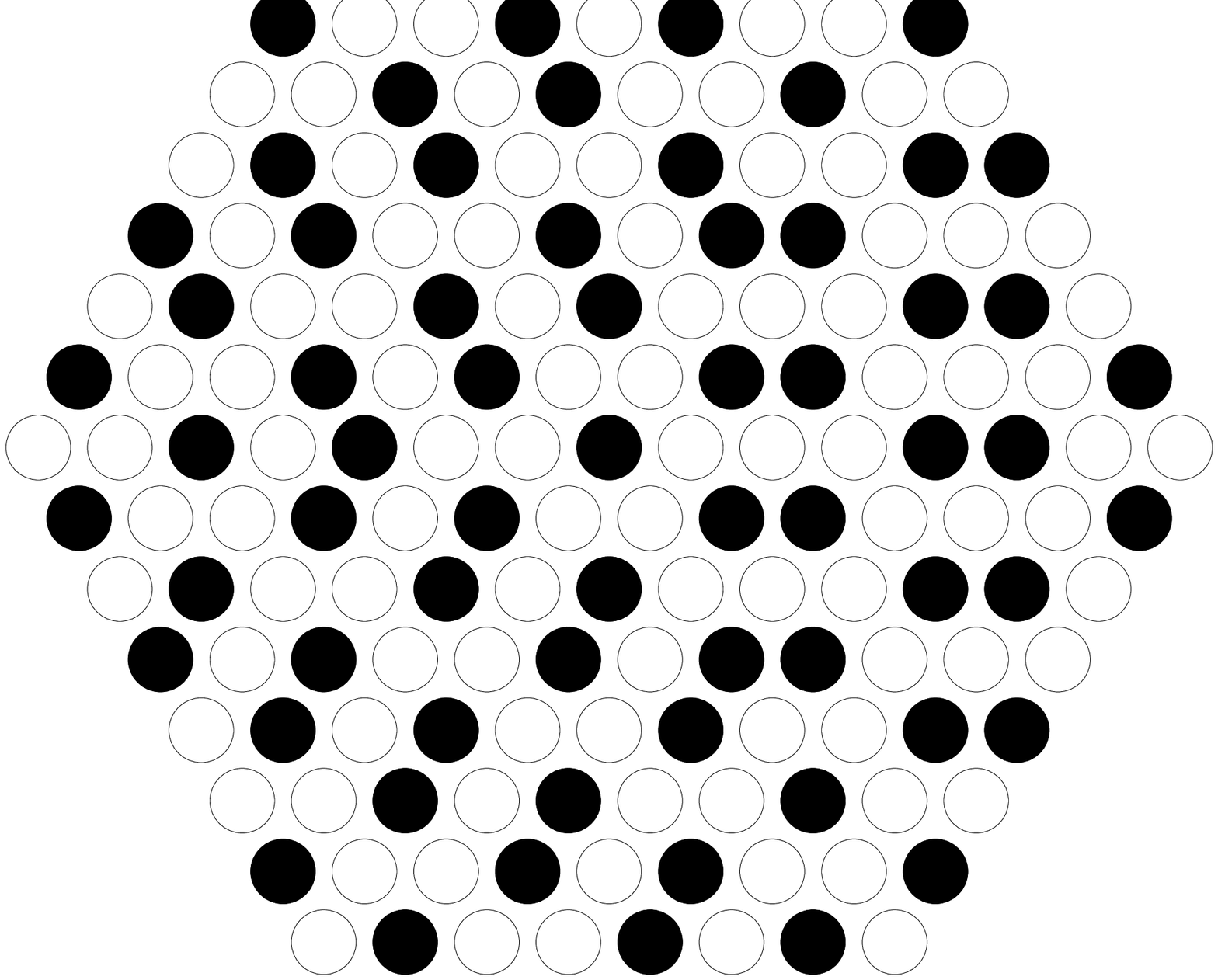}}\hfill \subfigure[\
$h = 2.0$ \label{f:D6z:hex1}] {\includegraphics*[bb = 30 340 565
840, width=\figwidths]{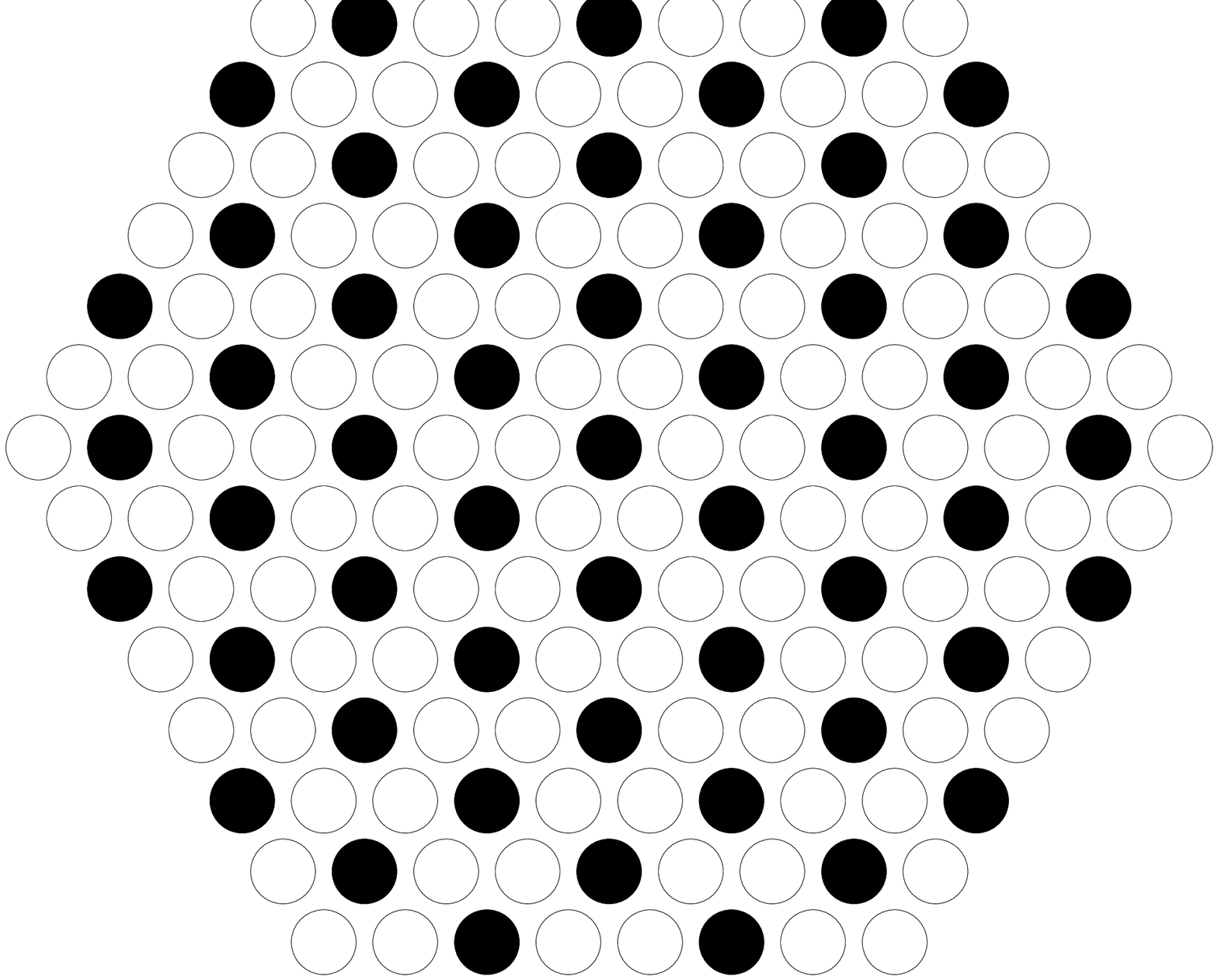}}\hfill \subfigure[\ $h = 6.0$
\label{f:D6z:hex2}] {\includegraphics*[bb = 30 340 565 840,
width=\figwidths]{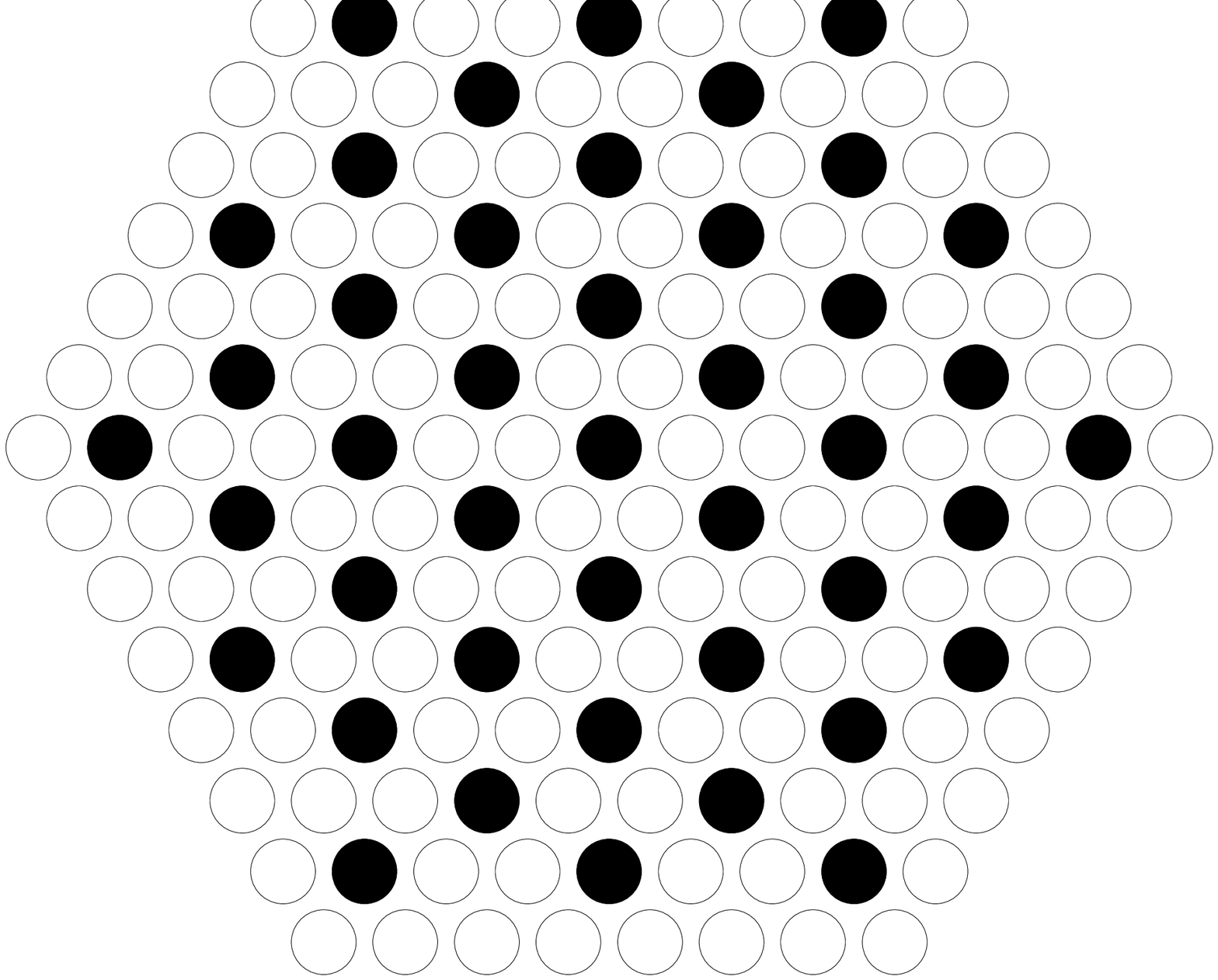}}\hfill \subfigure[\ $h = 6.1$
\label{f:D6z:hex3}] {\includegraphics*[bb = 30 340 565 840,
width=\figwidths]{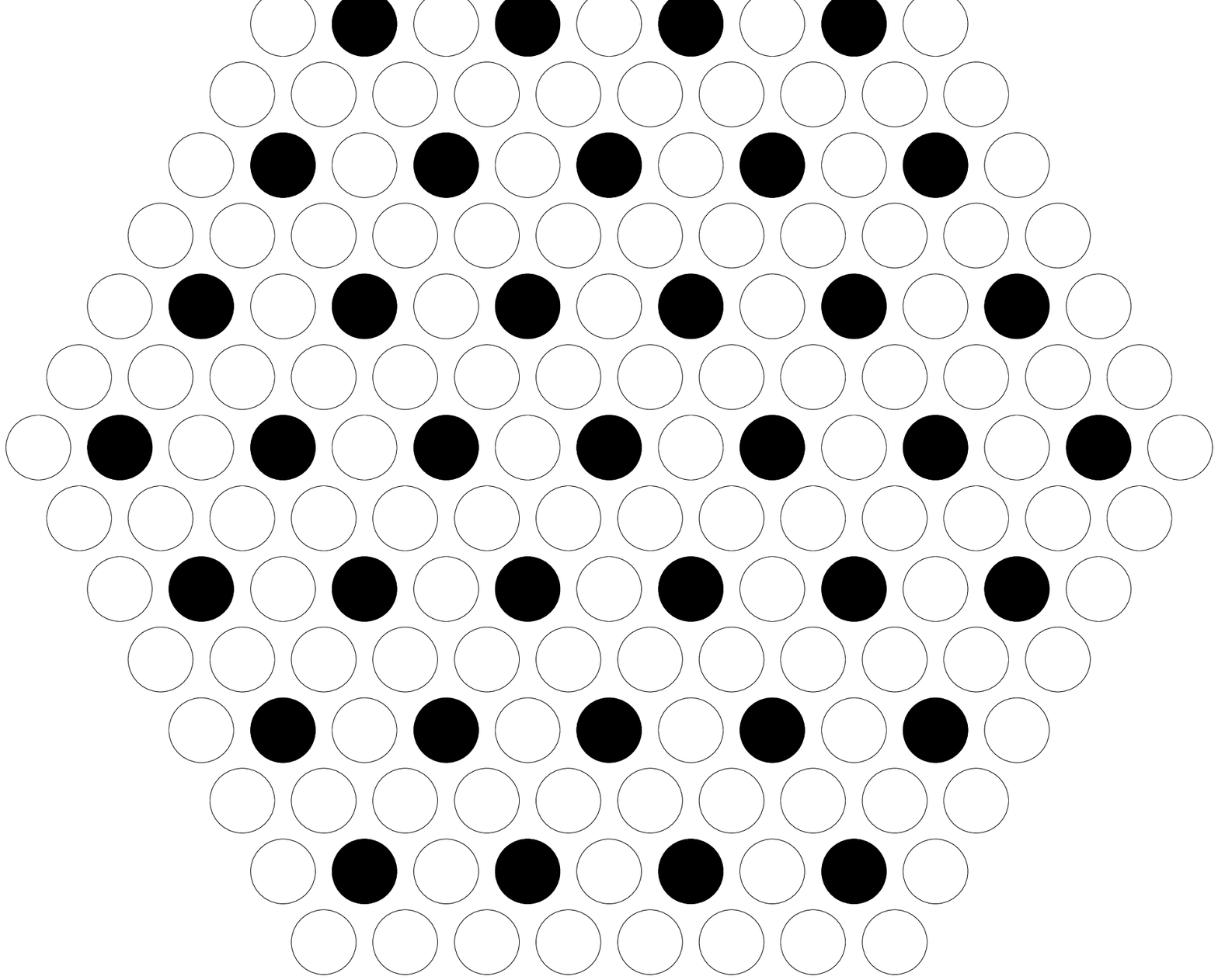}}\hfill \caption{Magnetic structure
for hexagonal sample with the size 7 at some characteristic values
of the magnetic field. \label{f:D6z:hex}}
\end{figure}

The magnetization function for the finite arrays in many details
repeats such a function for infinite system. All typical fields of
the structure change are reduced due to long-range nature of the
dipole-dipole interaction and its cropping on the length of the
order of the sample size $R$. This feature is more strikingly
expressed for the saturation field, where the respective sum
converges with the distance as $R^{-1}$, rather than for the field
of antiferromagnetic destruction, which evaluates by the rapid
convergent alternating sums. At small fields the same the
topological domain wall scenario is present, with the orientation of
the domain walls consistent with the shape of the sample, see
Fig.~\ref{f:D6z:hex0}. Step-like behavior (plateau behavior) of the
function of the infinite system corresponding to the becomes less
pronounced in the case of the finite array due to the
incommensurateness between the superlattice structure and the shape
of the sample. However the period is small for the triangular
superlattices of the small parameters ($\sqrt{3}$ и 2) and
incommensurateness does not have impact yet.  In this case typical
plateaus are visible on the magnetization curve and additional fine
structure of the form of the low steps appears on them. This fine
structure is relating to the change of superstructure contours but
not to its reorganization.

As has been shown in Sec.~\ref{semi-infinite}, the edges of finite
array are ``weak points'' for a single-dot instability, and the
values of the instability field for a dot at the vertex of edges are
lower than in the bulk, see Eq.~\eqref{eq.60} and
Eq.~\eqref{eq.120}. Thus for a finite sample with the increasing of
the external field the saturation starts at the border at lower
fields and the area of the superlattice will be reduced. Such
behavior is shown on the Fig.~\ref{f:D6z:hex1} и
Fig.~\ref{f:D6z:hex2}. As one can see from the Fig.~\ref{f:D6z:hex},
configurations for the hexagon are in the good correspondence with
the configuration for the infinite system. The only difference is
the presence of saturated dense border lines at high enough fields.

\section{DISCUSSION AND CONCLUSIONS}

Let us now discuss the general regularities revealed in the behavior
two two-dimensional systems with dipolar coupling of Ising magnetic
moments, the triangular lattice and the square lattice of magnetic
particles. First note that for nearest-neighbors interaction the
features of ordering for these two lattices are of principal
difference; the square lattice Ising moments at finite temperature
shows typical phase transition to the phase with the long range
chessboard AFM order, whereas the AFM Ising triangular lattice
system still unordered until $T\rightarrow 0$.\cite{GekhtObz,NoPT}
The reason, well discussed in the literature, is based on the
frustrated character of triangular magnets with AFM
interaction.\cite{GekhtObz} The nearest-neighbors Ising AFM system
is an extreme example of such feature, because it has topological
defects with zero energy, which can lead to destruction of
long-range order.\cite{Korshunov,KorshunovUFN}

For our case of long-ranged dipolar interaction the role of
frustration is not so crucial; it is enough to mention that for a
square lattice of Ising moments with dipolar coupling the
next-nearest-neighbors in AFM structure are frustrated as well. On
the other hand, the topological defect lines having zero energy for
triangular lattice nearest-neighbor system gain finite (but small)
energy for real dipolar interaction. But the effects of frustration
for nearest-neighbors interaction produce an essential difference in
the behavior of the systems of interest, magnetic dot arrays with
square and triangular lattices.

First let us mention the properties common for both square lattice
and triangular lattice. Their ground state at zero field is
antiferromagnetic, and the saturated (ferromagnetic) state is
present for high enough magnetic fields. For high magnetic fields,
nearly the saturation, the behavior of both lattices is practically
the same. The destruction of the saturated state for both lattices
happens through creation of a superlattice of reversed magnetic
moments, and at $h\to h_{sat}$ both superlattices are triangular. We
can only mention a quantitative difference of the saturation fields,
the value for triangular lattice, $11.034 m_0/a^3$, is a bit higher
than for a square lattice, $H_{\mathrm{sat,\ square}}=9.033622
m_0/a^3$. This difference just reflects the fact that the square
lattice is less dense than triangular. But for the region of low
fields, corresponding to the destruction of AFM state, and for
intermediate fields, where non-saturated states with $\langle
m\rangle =(0.2$-$0.8)m_0$ occurs, the behavior of these two lattices
is completely different.

For the square lattice the states with small, but non-zero  $\langle
m\rangle$ can be obtained from the chessboard AFM structure by
reversing the magnetic moments of a small fraction of magnetic dots
upward, leaving the remainder undisturbed.\cite{BishGalkIv} For a
triangular lattice, the destruction of the ideal AFM structure is
determined by creation of the system of topological lines (domain
walls). The creation of such a single line (or small density of such
lines) effects approximately one half of the particles in the
system. Thus, this phenomena appears at much lower field (compare
the values $2.646 m_0/a^3$ and $0.7086 m_0/a^3$ for these two
lattices) but needs to overcome much higher potential barrier.

For intermediate region of fields, the common property for both
lattices is that the ground states are mostly characterized by
complicated superlattices with different densities of ``up'' and
``down'' magnetic moments. But the process of magnetization for an
array with square dot lattice is influenced mainly by an interplay
between square and (distorted) triangular superlattices of ``down''
moments. Such behavior is dictated by a compromise between the
optimum of interaction energy (evident for triangular superlattice)
and the absence of the distortion energy for square superlattice
commensurate with the underlying lattice. In contrast, for a
triangular dot array triangular superlattices are optimal from any
point of view. Such structures occur at the main regions of field,
producing well-defined ``shelves''. However, the frustration effects
are present for any triangular lattices and superlattices, leading
to complicated character of the transition between the states with
superlattices of different period, especially small periods like
$a\sqrt{3}$ and $2a$.

It is also reasonable to discuss briefly problems that remain beyond
the scope of this paper. Of course, the problem of the transition
from the out-of-plane (Ising) states to planar states when the
magnetic field varies for finite anisotropy for a single dot is of
interest. Anisotropy energy for $n$-th dot can be written as
$(1/2)m_0 H_{\mathrm{an}} (\vec \sigma _{n}\cdot \vec {e_z})^2$,
where the anisotropy field, $H_{\mathrm{an}}$ characterizing the
strength of the anisotropy energy, is introduced. In fact, this
means the construction of a phase diagram on the $H, H_{\mathrm{an}}
$ plane. One can expect that Ising states are stable for
sufficiently large anisotropy, when the anisotropy energy is
comparable with the energy of interaction between neighboring dots,
$H_{\mathrm{an}} \geq m_0/a^3$. The analysis of the square lattice
corroborates this rule,\cite{BondGalkinIvJETP11} and demonstrates
that the Ising states are stable at $H_{\mathrm{an}} \geq  5
m_0/a^3$. It is rather difficult to analyze planar states
analytically. Our preliminary numerical data indicate that the
non-saturated states at small anisotropy correspond to complex
noncollinear structures, which are characterized both by significant
two-dimensional inhomogeneity with a scale of about the sample size,
and by the presence of regions where neighboring magnetic moments
are substantially noncollinear, and full description of such states
can be done only numerically. The complete analysis of non-collinear
states is evidently far beyond the scope of our article.

\section*{acknowledgments}
We are thankful to V.~G. Bar'yakhtar for useful discussions and
help. This work was partly supported by the National Ukrainian
Academy of Sciences via grant No.~228-11 and Scientific and
Technical Center of Ukraine (STCU), project No.~5210.

\section*{Appendix. Numerical method description}

To find the global minimum of the energy of the system the
Monte-Carlo approach with a simulated annealing (MC-SA) method was
used, see original articles Refs.~\onlinecite{MCSA1,MCSA2} and the
textbook Ref.~\onlinecite{MonteCarloBook}. For MC-SA realization,
first, the random initial configuration was selected. Every
iteration of the MC-SA method consists of $N$ moment reversal
attempts on the random site, where $N$  is the site number in the
sample. The main idea of simulated annealing is that the probability
of the reversal is non-zero even if the energy is growing after this
reversal; otherwise, the system with a high probability will be
``frozen''  in some local minimum. The probability depends not only
on energy gain for reversal, which equals to  $Hm_0$ ($H$ is a value
of the field created by other dots on a given ) but also on a global
time-varying parameter $T$ called the \emph{temperature}. If the
reversal is favorable in energy, the moment is always reversed,
irrespective of the temperature. But even if the reversal is
unfavorable, the non-zero probability of reversal is chosen as
follows: flip-over takes place if    $Hm_0 < T |\log p|$, where $T$
is the current value of temperature,   $p$  is a random value
generated in the range $0 < p \le 1$. Here the parameter temperature
determines the strategy of the minimization: for large $T$, the
evolution is sensitive to coarser energy variations, while it is
sensitive to finer energy variations when $T$ is small. Thus the
meaning of the temperature is the same as for annealing in
metallurgy involving initial heating and controlled cooling of a
material thereby avoiding defect formation.

The temperature is changing according to the quantity of full steps
of MC-SA of the sample $n$ as follows, $T= T_0 \min [\kappa, (n_0 /
n)^\alpha]$. Here the parameter $T_0$  was chosen as
(0.2-0.4)$m_0^2/a^3$, and the cutoff parameter $\kappa $ was equal
to $\kappa = 3.0$ such that the initial temperature was high enough
compared with the interaction energy. The optimal values of other
parameter are defined by the trial runs; $n_0$ was equal to  $ 10^4$
- $5\cdot 10^4$ and the value of the $\alpha$-index was taken 1/4 or
1/5.

The temperature decreases with the process evolution and
magnetization reversals take place more rarely with decreasing
energy. The process was stopped if the energy did not become less
than the previous minimum during the previous $10 n_0$ iterations.
Then the next random configuration was generated and the process was
repeated. Looking at the energies of minimal configurations for
every run one can estimate the probability that such a configuration
corresponds to the true minimum. For example, almost all processes
of the cooling go through the same minimum far from the regions of
the parameters which correspond to the transition between different
states.

For the configurations on the infinite lattice we took the rhombuses
with the consecutive increasing periods up to $16 \times 16$ and
used periodic boundary conditions. The energy of the configuration
was recalculated to the energy per one dot thereby making possible
the comparison of the results for the different periods. A rhombus
was selected to admit the maximum possible configuration set to
consideration, but minimal configurations often have higher
symmetry.

A similar approach was used to build a magnetization function.
Firstly, all ranges of the fields were passed with large intervals
and small periods of the rhombuses. After that the regions of the
structure rebuilding were detected, where the field intervals were
taken smaller and periods were taken larger.

\end{document}